\documentclass[aps,pre,superscriptaddress,12pt]{revtex4}

\usepackage{graphicx,epic,eepic,epsfig,amsmath,latexsym,amssymb,verbatim}

\usepackage{theorem}
\newtheorem{definition}{Definition}
\newtheorem{lemma}[definition]{Lemma}
\newtheorem{theorem}[definition]{Theorem}
\newtheorem{corollary}[definition]{Corollary}

\def\squareforqed{\hbox{\rlap{$\sqcap$}$\sqcup$}}
\def\qed{\ifmmode\squareforqed\else{\unskip\nobreak\hfil
\penalty50\hskip1em\null\nobreak\hfil\squareforqed
\parfillskip=0pt\finalhyphendemerits=0\endgraf}\fi}
\def\endenv{\ifmmode\;\else{\unskip\nobreak\hfil
\penalty50\hskip1em\null\nobreak\hfil\;
\parfillskip=0pt\finalhyphendemerits=0\endgraf}\fi}

\mathchardef\ordinarycolon\mathcode`\: \mathcode`\:=\string"8000
\def\vcentcolon{\mathrel{\mathop\ordinarycolon}}
\begingroup \catcode`\:=\active
  \lowercase{\endgroup
  \let :\vcentcolon
  }

\newcommand{\id}{{\sf 1 \hspace{-0.3ex} \rule{0.1ex}{1.52ex}\rule[-.01ex]{0.3ex}{0.1ex}}}

\begin{document}
\title{Unambiguous discrimination among oracle operators}
\author{Anthony Chefles}
\email{anthony.chefles@hp.com}
\affiliation{Quantum Information
Processing Group, Hewlett-Packard Laboratories, Filton Road, Stoke
Gifford, Bristol BS34 8QZ, U.K.}
\author{Akira Kitagawa}
\affiliation{National Institute of Information and Communications
Technology (NICT), 4-2-1 Nukui-Kita, Koganei, Tokyo 184-8795, Japan}
\affiliation{Core Research for Evolutional Science and Technology
(CREST),
Japan Science and Technology Agency, \\
1-9-9 Yaesu, Chuoh, Tokyo 103-0028, Japan} \affiliation{ Department
of Photonics, Faculty of Science and Engineering, Ritsumeikan
University, 1-1-1, Noji-Higashi, Kusatsu City, Shiga 525-8577,
Japan}
\author{Masahiro Takeoka}
\affiliation{National Institute of Information and Communications
Technology (NICT), 4-2-1 Nukui-Kita, Koganei, Tokyo 184-8795, Japan}
\affiliation{Core Research for Evolutional Science and Technology
(CREST),
Japan Science and Technology Agency, \\
1-9-9 Yaesu, Chuoh, Tokyo 103-0028, Japan}
\author{Masahide Sasaki}
\affiliation{National Institute of Information and Communications
Technology (NICT), 4-2-1 Nukui-Kita, Koganei, Tokyo 184-8795, Japan}
\affiliation{Core Research for Evolutional Science and Technology
(CREST),
Japan Science and Technology Agency, \\
1-9-9 Yaesu, Chuoh, Tokyo 103-0028, Japan}
\author{Jason Twamley}
\affiliation{Centre for Quantum Computer Technology, Macquarie
University, Sydney, New South Wales 2109, Australia}
\begin{abstract}
\vspace{0.5cm}We address the problem of unambiguous discrimination
among oracle operators. The general theory of unambiguous
discrimination among unitary operators is extended with this
application in mind. We prove that entanglement with an ancilla
cannot assist any discrimination strategy for commuting unitary
operators.  We also obtain a simple, practical test for the
unambiguous distinguishability of an arbitrary set of unitary
operators on a given system.  Using this result, we prove that the
unambiguous distinguishability criterion is the same for both
standard and minimal oracle operators. We then show that, except in
certain trivial cases, unambiguous discrimination among all standard
oracle operators corresponding to integer functions with fixed
domain and range is impossible. However, we find that it is possible
to unambiguously discriminate among the Grover oracle operators
corresponding to an arbitrarily large unsorted database. The
unambiguous distinguishability of standard oracle operators
corresponding to totally indistinguishable functions, which possess
a strong form of classical indistinguishability, is analysed. We
prove that these operators are not unambiguously distinguishable for
any finite set of totally indistinguishable functions on a Boolean
domain and with arbitrary fixed range. Sets of such functions on a
larger domain can have unambiguously distinguishable standard oracle
operators and we provide a complete analysis of the simplest case,
that of four functions. We also examine the possibility of
unambiguous oracle operator discrimination with multiple parallel
calls and investigate an intriguing unitary superoperator
transformation between standard and entanglement-assisted minimal
oracle operators.
\end{abstract}
\maketitle
\newpage

\section{Introduction}
\renewcommand{\theequation}{1.\arabic{equation}}
\setcounter{equation}{0} \label{sec:1}

One of the most important problems in theoretical computer science
is the oracle identification problem.  This can be described in the
following way. We are given a device, known as an oracle, which is
promised to compute one of a known set of functions. The oracle
identification problem consists of determining which function the
oracle computes. It is understood that we are not permitted to
investigate the internal workings of the device. Instead, it is
treated as a black box. The only information at our disposal is our
record of the input and of the output it gives rise to.

The action of the oracle is a physical process by which the output
is computed from the input.  To the best of our current knowledge,
all physical processes are quantum mechanical. If we are to describe
the action of the oracle quantum mechanically, it will be
represented by a quantum channel, or operation, with a different
operation corresponding to each possible function. However, quantum
channels, unitary channels in particular, can operate coherently on
superpositions of quantum states, giving rise to the well-known
phenomenon of quantum parallelism.  This parallelism can be
exploited to perform oracle identification with lower query
complexity, i.e. with fewer uses of the oracle, than can be achieved
classically.

As a consequence of the coherent information processing capabilities
of unitary operators, in quantum computation, oracles are
conventionally taken to be unitary processes.  The quantum oracle
identification problem is then essentially a problem of
discrimination among individual, or sets of unitary operators, where
each operator coherently computes one of a known set of functions.
These operators are naturally known as oracle operators. It is often
unnecessary to distinguish among all of the possible oracle
operators corresponding to a given set of functions individually,
but only among subsets of the total possible set, for the advantages
of a quantum over a classical oracle to become evident. Indeed, the
first demonstrations of quantum computational speed-up, those
apparent in the Deutsch \cite{Deutsch} and later Deutsch-Jozsa
\cite{DJ} algorithms, which demonstrated accelerated discrimination
between uniform and balanced functions, exemplified the enhanced
distinguishability of sets of quantum oracle operators over their
classical counterparts. Discrimination among sets of functions with
different periodicities is central to the Simon \cite{Simon} and
Shor \cite{Shor} algorithms. Again, it is the fact that this can be
carried out more efficiently with quantum oracle operators than
known classical methods which is responsible for the quantum
computational speed-up. The quantum searching algorithm discovered
by Grover \cite{Grover} can also be interpreted as an oracle
identification problem \cite{NC}, although one where the aim is a
fine-grained discrimination among individual functions rather than
larger sets.

It was developments such as these, which demonstrated the superior
distinguishability of quantum oracle operators over corresponding
classical channels for specific classes of functions, that led to
the oracle identification problem being investigated in general
terms. If we wish to identify an unknown function of an $M$-ary
variable, then classically, we must evaluate it for each of these
possible values, which implies $M$ oracle calls. Quantum
mechanically, however, it was shown by van Dam \cite{vandam} that a
quantum oracle corresponding to such a function can be identified
with probability $>0.95$ with $M/2+\mathrm{O}(\sqrt{M})$ calls.
Further general results relating to quantum query complexity have
been obtained by Iwama and collaborators \cite{AIKMPY,IKRS} and by
Fahri et al. \cite{FGGS}. In particular, the latter authors
established an upper bound on the number of functions that can be
identified with a fixed number of calls and a given correct
identification probability. Here, the different functions were
encoded in generally non-orthogonal states and distinguished using a
projective measurement, with each  outcome corresponding
conclusively to one of the possible functions.  The impossibility of
perfect discrimination among non-orthogonal states with a projective
measurement implies that there would be a non-zero probability of
this result being erroneous.

However, it is sometimes possible to distinguish among
non-orthogonal states using the alternative strategy of unambiguous
state discrimination \cite{discreview1,discreview2}. Here, we are
not always guaranteed a conclusive result, although when one is
obtained, it will necessarily be correct.  As such, unambiguous
state discrimination is inherently probabilistic. Unlike state
discrimination strategies where we tolerate errors, it is not
possible to unambiguously discriminate among an arbitrary set of
states. For a set of pure states to be unambiguously
distinguishable, they must be linearly independent \cite{C1} and a
more complex constraint applies to general mixed states
\cite{CLOCC}.

In relation to oracle identification, the potential applicability of
unambiguous discrimination was first investigated by Bergou et al.
\cite{BHH,BH}. These authors demonstrated how one can obtain
generalisations of the Deutsch-Jozsa algorithm, where the oracle
operators encode information in non-orthogonal states.  They
nevertheless yield unambiguously correct information. This suggests
that unambiguous discrimination may have an important role to play
in quantum information processing, particularly in relation to
probabilistic algorithms.

The purpose of this article is to explore this possibility further.
We address the problem of unambiguous discrimination among oracle
operators in general.  In order to investigate this matter, it is
helpful to have a broad understanding of unambiguous discrimination
among unitary operators \cite{CS}.  As such, Section \ref{sec:2} is
devoted to presenting some preliminaries, some of which are new
results, relating to this general problem. Among these is a simple,
practical criterion for determining when an arbitrary set of unitary
operators is unambiguously distinguishable and a proof that
entanglement with an ancilla cannot aid any discrimination or
estimation procedure for commuting unitary operators. Section
\ref{sec:3} is concerned with another preliminary topic, the
properties of oracle operators. Here, we describe the main
properties of standard oracle operators, which can be constructed
for all functions from $\mathbb{Z}_{M}{\mapsto}\mathbb{Z}_{N}$ with
arbitrary positive integers $M,N$ \cite{ring} and minimal oracle
operators, which have the advantage of acting on a smaller register
although they are possible only for invertible functions
\cite{KKVB}. Throughout, we take these to be permutations.
 In contrast with other treatments, we make novel use of the
Pegg-Barnett phase operator \cite{PB}, as we find that this can be
used to obtain an appealing and useful compact representation of
standard oracle operators. In Section \ref{sec:4}, we apply our
general criterion for the unambiguous distinguishability of unitary
operators to both standard and minimal oracle operators. Remarkably,
it is found that the unambiguous distinguishability criterion is the
same for both kinds of oracle operator.

The next two sections are concerned with applying this criterion to
oracle operators corresponding to various interesting sets of
functions. In Section \ref{sec:5}, we show that it is impossible to
unambiguously discriminate among the standard oracle operators
corresponding to all functions from
$\mathbb{Z}_{M}{\mapsto}\mathbb{Z}_{N}$ for any fixed $M$ and $N$
both ${\geq}2$. However, we also show that the Grover oracle
operators corresponding to an arbitrary sized unsorted database can
be unambiguously discriminated with one shot.  This is noteworthy
because perfect discrimination among Grover oracle operators is
possible only for an unsorted database with at most four entries
\cite{BBHT}.

Section \ref{sec:6} is concerned with oracle operators corresponding
to sets of functions which we refer to as being totally
indistinguishable. A totally indistinguishable set of functions is a
set for which one can never determine which function was computed by
a classical oracle with known input and output data.  It is found
that, if the functions are distinct, then there must be at least
four functions in a set with this property. We analyse in some
detail the situation where the input variable is Boolean.  It is
found that sets of such functions admit a simple graphical
representation in terms of which the total indistinguishability
condition takes a geometrically appealing form. This representation,
together with various graph-theoretic results which apply to it, is
used to prove that for no finite set of totally indistinguishable
functions from $\mathbb{Z}_{2}{\mapsto}\mathbb{Z}_{N}$ are the
corresponding standard oracle operators unambiguously
distinguishable for any integer $N{\geq}2$.  This is not the case
for totally indistinguishable functions on a larger domain. We
present a complete characterisation of sets of four functions whose
domain is at least three-valued and with arbitrary, fixed finite
integer range $N{\geq}2$ which are totally indistinguishable yet
whose standard oracle operators are unambiguously distinguishable.

In Section \ref{sec:7}, we consider the possibility of unambiguous
oracle operator discrimination with multiple calls.  In this
article, we focus mainly on unambiguous discrimination among oracle
operators with just one call to the oracle.  If this is not
possible, the oracle operators may nevertheless be unambiguously
distinguishable if we are permitted $C>1$ calls.  We restrict our
attention to parallel calls, where registers are not reused for
subsequent calls.  We obtain sufficient conditions, in terms of
properties of the set of functions in question, for this to be
possible.

Section \ref{sec:8} is devoted to discussing the relationship
between standard and entanglement-assisted minimal oracle operators.
It is found that they have an intriguing unitary superoperator
interconvertibility property, whose implications are explored.  We
finally conclude in Section \ref{sec:9} with a discussion of our
results and suggestions for future research on this topic.

\section{Unambiguous discrimination among unitary operators}
\renewcommand{\theequation}{2.\arabic{equation}}
\setcounter{equation}{0} \label{sec:2}

To address the problem of unambiguous discrimination among oracle
operators, it will be helpful to have an appreciation of what can be
achieved in relation to unambiguous discrimination among unitary
operators in general and of any general limitations that apply.
Discrimination among a set of unitary operators is achieved be
letting one of them act upon an initial probe statend and then
discriminating among the possible output states in order to
determine which operator was implemented. The most general scenario
we can consider is the following.  Imagine that we have a quantum
system $Q$ with $D_{Q}$-dimensional Hilbert space ${\cal H}_{Q}$.
Suppose that there is also an ancilla $A$ having $D_{A}$-dimensional
Hilbert space ${\cal H}_{A}$, where $D_{Q}{\leq}D_{A}$.  These two
systems are initially prepared in a joint, possibly entangled probe
state.  We may take this initial state to be pure by considering a
sufficiently large ancilla, which we will do and write this state as
$|{\psi}_{QA}{\rangle}{\in}{\cal H}_{QA}={\cal H}_{Q}{\otimes}{\cal
H}_{A}$.   We then act on $Q$ with one of $K$ unitary operators
$U_{j}$, where $j=1,{\ldots},K$. The $K$ possible final states after
this action will be denoted by $|{\psi}_{QAj}{\rangle}$.   Our task
is to determine which of these states was produced, which will in
turn tell us which of the $U_{j}$ acted on $Q$.

To do so unambiguously, that is, with zero probability of error but
allowing for some probability (strictly) $<1$ of an inconclusive
result for each $j$, we require the $|{\psi}_{QAj}{\rangle}$ to be
linearly independent \cite{C1}. We may then ask: what properties
must the $U_{j}$ possess to produce a linearly independent set of
output states for at least one possible probe state
$|{\psi}_{QA}{\rangle}$, since this is clearly the condition for the
$U_{j}$ being unambiguously distinguishable. It is known that:

\begin{theorem}
\label{thmlind}
A necessary and sufficient condition for $K$ unitary
operators acting on a Hilbert space ${\cal H}_{Q}$ to be
unambiguously distinguishable is that they are linearly independent.
Moreover, a linearly independent set of unitary operators can always
be unambiguously discriminated using any probe state with maximum
Schmidt rank $D_{Q}$, the dimensionality of ${\cal H}_{Q}$.
\end{theorem}

This was proven originally by Chefles and Sasaki
 as a special case of a more general result \cite{CS}.  However, for
the sake of both completeness and convenience, we provide here a
simplified proof.\\

\noindent{\bf Proof}: We begin by proving, by contradiction, the
necessity of the linear independence of the unitary operators
$U_{j}$ for them to be amenable to unambiguous discrimination. If
these operators are linearly dependent, then there exist $K$
coefficients $a_{j}$, not all of which are zero, such that
$\sum_{j=1}^{K}a_{j}U_{j}=0$. For these coefficients, we have, for
any probe state $|{\psi}_{QA}{\rangle}$,
\begin{equation}
\label{lindep}
\sum_{j=1}^{K}a_{j}|{\psi}_{QAj}{\rangle}=\left[\left(\sum_{j=1}^{K}a_{j}U_{j}\right){\otimes}{\id}_{A}\right]|{\psi}_{QA}{\rangle}=0,
\end{equation}
where ${\id}_{A}$ is the identity operator on the ancilla Hilbert
space ${\cal H}_{A}$.  This shows that the final states
$|{\psi}_{QAj}{\rangle}$ are linearly dependent for any probe state
$|{\psi}_{QA}{\rangle}$ and are thus unamenable to unambiguous
discrimination.  It follows that linearly dependent unitary
operators cannot be unambiguously discriminated.

To prove that the $U_{j}$ can be always be unambiguously
discriminated if they are linearly independent and that this can be
achieved with any probe state $|{\psi}_{QA}{\rangle}$ which has
maximum Schmidt rank, we again use an argument by contradiction.  We
will employ the following:

\begin{lemma}
\label{lma:zerop} Let $|{\psi}_{QA}{\rangle}$, where
$D_{Q}{\leq}D_{A}$, be a state vector in ${\cal H}_{QA}$ with
maximum Schmidt rank, i.e. Schmidt rank=$D_{Q}$. The only operator
$H$ acting on ${\cal H}_{Q}$ for which
$(H{\otimes}{\id}_{A})|{\psi}_{QA}{\rangle}=0$ is the zero operator.
\end{lemma}

\noindent{\bf Proof}:  Let us write $|{\psi}_{QA}{\rangle}$ in
Schmidt decomposition form:
\begin{equation}
|{\psi}_{QA}{\rangle}=\sum_{k=1}^{D_{Q}}c_{k}|r_{k}{\rangle}{\otimes}|s_{k}{\rangle},
\end{equation}
where the $|r_{k}{\rangle}$ form an orthonormal basis for ${\cal
H}_{Q}$, the $|s_{k}{\rangle}$ form an orthonormal subset of ${\cal
H}_{A}$ and the Schmidt coefficients $c_{k}$ are, by assumption, all
non-zero. If $(H{\otimes}{\id}_{A})|{\psi}_{QA}{\rangle}=0$, then
upon taking the inner product throughout on the $A$ system with an
arbitrary element of the set $\{|s_{k}{\rangle}\}$ and making use of
the fact that the corresponding Schmidt coefficient is non-zero, we
find that $H|r_{k}{\rangle}=0$ for all $k$.  Hence all matrix
elements of $H$ in the $|r_{k}{\rangle}$ basis are zero and so $H$
is the zero operator.  This completes the proof. $\Box$\\

To make use of this, suppose that we have a probe state
$|{\psi}_{QA}{\rangle}$ with maximum Schmidt rank such that the
final states $|{\psi}_{QAj}{\rangle}$ are unamenable to unambiguous
discrimination, i.e. they are linearly dependent.  Then there exist
coefficients $a_{j}$, not all of which are zero, such that Eq.
\eqref{lindep} is true. Applying Lemma \ref{lma:zerop}, with
$H=\sum_{j=1}^{K}a_{j}U_{j}$, we see that the $U_{j}$ must be
linearly
dependent.  This completes the proof $\Box$.\\

An interesting question is whether or not a probe state with maximum
Schmidt rank can always be used for optimum unambiguous unitary
operator discrimination, i.e. for attaining the theoretical minimum
probability of inconclusive results.  At this time, the answer to
this question is unknown.

We see that to unambiguously discriminate among a set of unitary
operators, we require them to be linearly independent. This prompts
us to ask if there is a simple, practical test for the linear
independence of a set of unitary operators.  Since unitary operators
are vectors in a vector space, it is natural to imagine that a
general test for the linear independence of vectors can be easily
adapted to operators.  This is indeed the case.  Consider, for
example, a $D$-dimensional vector space ${\cal V}$ endowed with
inner product ${\langle}u,w{\rangle}=\sum_{k=1}^{D}u^{*}_{k}w_{k}$,
where $u=(u_{k})$ and $w=(w_{k})$ are two arbitrary vectors in
${\cal V}$ represented in some common orthonormal basis.  The linear
independence of an arbitrary set of vectors
$u_{j}=(u_{jk}){\in}{\cal V}$ can be checked by calculating the
positive semi-definite Gram matrix
\begin{equation}
G=(G_{j'j})=({\langle}u_{j'},u_{j}{\rangle}).
\end{equation}
It is a well-known result from elementary linear algebra that the
$u_{j}$ are linearly independent iff $G$ is non-singular.

To apply this to a set of $K$ unitary operators $U_{j}$ acting on
${\cal H}_{Q}$, we simply rearrange their components in some fixed
basis as the components of corresponding vectors in
$\mathbb{C}^{D_{Q}^{2}}$, having the above inner product.  We then
find that the inner product of such `vectorisations' of two
unitaries $U$ and $W$ is simply $\mathrm{Tr}(U^{\dagger}W)$.  Hence,
to determine whether or not the unitary operators $U_{j}$ are
linearly independent, we calculate their Gram matrix $G=(G_{j'j})$,
whose elements are
\begin{equation}
\label{Grammatrix} G_{j'j}=\mathrm{Tr}(U^{\dagger}_{j'}U_{j}).
\end{equation}
Our condition for the linear independence of these operators is
simply that
\begin{equation}
\det(G)>0.
\end{equation}
We shall make extensive use of this condition in subsequent
sections.

We mention in passing that the Gram matrix determinant (the
`Grammian') of a set of quantum states plays an important role in
unambiguous quantum state comparison \cite{CAJ}.  If we have a set
of similar quantum systems all prepared in unknown pure states, then
one can unambiguously confirm that these states are all different
iff their Grammian is non-zero.  Indeed, the statistics of the
optimum measurement for confirming this, which separates the
antisymmetric and non-antisymmetric subspaces of the systems,
directly measure the Grammian.

In this article, we will be concerned with unambiguous
discrimination among quantum oracle operators. As we shall see in
the next section, the standard quantum oracle operators
corresponding to functions with fixed domain and range form an
Abelian group and thus mutually commute. The following theorem shows
that initial entanglement cannot help us discriminate among
commuting unitary operators:\\

\begin{theorem}
\label{commute} If a set of unitary operators $U_{j}$ mutually
commute, then for any possibly entangled probe state
$|{\psi}_{QA}{\rangle}{\in}{\cal H}_{QA}$, one can produce the
corresponding output states
$|{\psi}_{QAj}{\rangle}=(U_{j}{\otimes}{\id}_{A})|{\psi}_{QA}{\rangle}$
in an alternative manner by preparing the systems $Q$ and $A$
initially in a product probe state $|{\xi}_{QA}{\rangle}$, following
which one of the $U_{j}$ acts on $Q$ and then finally $Q$ and $A$
interact via some unitary
operator $V$ acting on ${\cal H}_{QA}$ which is independent of $j$. \\
\end{theorem}

\noindent{\bf Proof}: Consider a set of unitary operators $U_{j}$
acting on ${\cal H}_{Q}$.  If these commute, then they can be
simultaneously diagonalised and therefore written as
\begin{equation}
U_{j}=\sum_{k=1}^{D_{Q}}e^{i{\omega}_{jk}}|{\alpha}_{k}{\rangle}{\langle}{\alpha}_{k}|,
\end{equation}
where the ${\omega}_{jk}$ are real and the set
$\{|{\alpha}_{k}{\rangle}\}$ is an orthonormal basis for ${\cal
H}_{Q}$.  The systems $Q$ and $A$ are initially prepared in the
probe state $|{\psi}_{QA}{\rangle}$, which may be entangled.  We can
write this state as
\begin{equation}
\label{schmidt}
|{\psi}_{QA}{\rangle}=\sum_{k=1}^{D_{Q}}\sum_{l=1}^{D_{A}}c_{kl}|{\alpha}_{k}{\rangle}{\otimes}|{\beta}_{l}{\rangle},
\end{equation}
where the set $\{|{\beta}_{l}{\rangle}\}$ is an orthonormal basis
set for ${\cal H}_{A}$.  The coefficients $c_{kl}$ satisfy
$\sum_{k=1}^{D_{Q}}\sum_{l=1}^{D_{A}}|c_{kl}|^{2}=1$.  The final
states $|{\psi}_{QAj}{\rangle}$ are obtained through
\begin{equation}
|{\psi}_{QAj}{\rangle}=(U_{j}{\otimes}{\id}_{A})|{\psi}_{QA}{\rangle}=\sum_{k=1}^{D_{Q}}\sum_{l=1}^{D_{A}}c_{kl}e^{i{\omega}_{jk}}|{\alpha}_{k}{\rangle}{\otimes}|{\beta}_{l}{\rangle},
\end{equation}
where ${\id}_{A}$ is the identity operator on ${\cal H}_{A}$.
Crucially, the Gram matrix of this set of states has the elements
\begin{equation}
\label{entgram}
{\langle}{\psi}_{QAj'}|{\psi}_{QAj}{\rangle}=\sum_{k=1}^{D_{Q}}\sum_{l=1}^{D_{A}}|c_{kl}|^{2}e^{i({\omega}_{jk}-{\omega}_{j'k})}=\sum_{k=1}^{D_{Q}}p_{k}e^{i({\omega}_{jk}-{\omega}_{j'k})}.
\end{equation}
Here, we have defined
\begin{equation}
\label{pk} p_{k}=\sum_{l=1}^{D_{A}}|c_{kl}|^{2}.
\end{equation}
Clearly, we have $p_{k}{\geq}0$ and $\sum_{k=1}^{D_{Q}}p_{k}=1$. For
any $p_{k}$ satisfying these two conditions, consider instead
preparing the initial product state
\begin{equation}
|{\xi}_{QA}{\rangle}=|{\xi}{\rangle}{\otimes}|{\chi}{\rangle},
\end{equation}
where
\begin{equation}
\label{qinit}
|{\xi}{\rangle}=\sum_{k=1}^{D_{Q}}\sqrt{p_{k}}|{\alpha}_{k}{\rangle}
\end{equation}
and $|{\chi}{\rangle}$ is an arbitrary normalised state in ${\cal
H}_{A}$.  Suppose that we had started with this state rather than
the state $|{\psi}_{QA}{\rangle}$.  Then, upon application of
$U_{j}$, we would have obtained
\begin{equation}
|{\xi}_{QAj}{\rangle}=(U_{j}|{\xi}{\rangle}){\otimes}|{\chi}{\rangle}=|{\xi}_{j}{\rangle}{\otimes}|{\chi}{\rangle},
\end{equation}
where
\begin{equation}
|{\xi}_{j}{\rangle}=\sum_{k=1}^{D_{Q}}\sqrt{p_{k}}e^{i{\omega}_{jk}}|{\alpha}_{k}{\rangle}.
\end{equation}
Let us now calculate the elements of the Gram matrix of these
states. We obtain
\begin{equation}
{\langle}{\xi}_{QAj'}|{\xi}_{QAj}{\rangle}={\langle}{\xi}_{j'}|{\xi}_{j}{\rangle}=\sum_{k=1}^{D_{Q}}p_{k}e^{i({\omega}_{jk}-{\omega}_{j'k})},
\end{equation}
which are equal to the Gram matrix elements in Eq. \eqref{entgram}
for the possibly entangled probe state $|{\psi}_{QA}{\rangle}$. Two
sets of states with the same Gram matrix can be unitarily
transformed into each other \cite{DG,JS}. We could therefore begin
with the non-entangled probe state $|{\xi}_{QA}{\rangle}$, let one
of the $U_{j}$ act, then perform a single unitary transformation $V$
on $QA$ to get the final state $|{\psi}_{QAj}{\rangle}$ which we
would have obtained had we started with the potentially entangled
probe state $|{\psi}_{QA}{\rangle}$, for all $j$. This proves that
entanglement with an ancilla gives no advantage in
attempting to discriminate among commuting unitary operators $\Box$. \\

Notice that the applicability of the above result is not limited to
unambiguous discrimination.   It applies to every unitary operator
discrimination strategy including, e.g. minimum error
discrimination.  It also applies to estimation strategies where the
index $j$ labels the possible values of one or more parameters to be
estimated. Indeed, if necessary, it is a straightforward matter to
replace $j$ with one or more continuous indices. Doing so provides
an alternative method of arriving at the main conclusions of
\cite{MAB}.

The above theorem has the following consequences:

\begin{corollary}
If $K$ commuting unitary operators acting on a $D_{Q}$-dimensional
Hilbert space can be unambiguously discriminated, then
\begin{equation}
\label{kd} K{\leq}D_{Q}.
\end{equation}
\end{corollary}
\noindent{\bf Proof}: This is a simple consequence of the fact that
we can always neglect the ancilla $A$ and concentrate on $Q$ and
that at most
$D_{Q}$ states in ${\cal H}_{Q}$ can be linearly independent and therefore unambiguously discriminated $\Box$.\\

\begin{corollary}
If $K$ commuting unitary operators $U_{j}$ acting on a
$D_{Q}$-dimensional Hilbert space can be unambiguously
discriminated, then this can always be achieved using any probe
state in ${\cal H}_{Q}$ which is a maximal superposition of the
(common) eigenstates of the $U_{j}$.
\end{corollary}
\noindent{\bf Proof}: Any such state of $Q$ can be written as
$|{\xi}{\rangle}=\sum_{k}\sqrt{p_{k}}e^{i{\theta}_{k}}|{\alpha}_{k}{\rangle}$
for some angles ${\theta}_{k}$ and where $p_{k}>0$.  Then the states
$(U_{j}{\otimes}{\id}_{A})|{\xi}{\rangle}{\otimes}|{\chi}{\rangle}$,
for any pure state $|{\chi}{\rangle}{\in}{\cal H}_{A}$, have the
same Gram matrix as $(U_{j}{\otimes}{\id}_{A})|{\psi}_{QA}{\rangle}$
where
$|{\psi}_{QA}{\rangle}=\sum_{k}\sqrt{p_{k}}e^{i{{\theta}_{k}}}|{\alpha}_{k}{\rangle}{\otimes}|{\alpha}_{k}{\rangle}$,
which has maximum Schmidt rank.  The corollary follows from the
equality of these Gram matrices (and thus the unitary
interconvertibility of these sets of states) and Theorem \ref{thmlind} $\Box$.\\

Theorem \ref{commute} tells us that the set of states produced with
a possibly entangled probe state $|{\psi}_{QA}{\rangle}$ can always
be produced with an unentangled probe state $|{\xi}_{QA}{\rangle}$
and postprocessing with some bipartite unitary operator $V$,
implying that entanglement gives no advantage in any strategy for
discrimination among commuting unitary operators. From this, we can
see that the ancilla and the bipartite unitary interaction $V$ can
be removed altogether from the preparation procedure and absorbed
into the ancilla and interaction involved in the (generalised)
measurement used to discriminate (for any strategy) among the
monopartite states $|{\xi}_{j}{\rangle}$, where all the information
about which operator was applied is contained. As such, in what
follows, whenever discussing the preparation aspects of
discrimination among commuting unitary operators, unless stated
otherwise, we shall no longer assume there to be an ancilla $A$.

\section{Properties of oracle operators}
\renewcommand{\theequation}{3.\arabic{equation}}
\setcounter{equation}{0}
\label{sec:3}

\subsection{Standard oracle operators}

Let $M,N$ be arbitrary integers ${\geq}1$.  Consider ${\cal
F}_{MN}$, the set of functions from
$\mathbb{Z}_{M}{\mapsto}\mathbb{Z}_{N}$.    We take $M,N<{\infty}$
throughout this article except in one specific situation that we
discuss in Section \ref{sec:6}, which will be clear when it arises.
Let ${\cal H}_{M}$ and ${\cal H}_{N}$ be $M$- and $N$-dimensional
Hilbert spaces respectively. To each $f{\in}{\cal F}_{MN}$ there
corresponds a unitary standard oracle operator on ${\cal
H}_{M}{\otimes}{\cal H}_{N}$:
\begin{equation}
\label{oracleaction}
U_{f}|x{\rangle}{\otimes}|y{\rangle}=|x{\rangle}{\otimes}|y{\oplus}f(x){\rangle}.
\end{equation}
Here, ${\oplus}$ denotes addition modulo $N$. Also,
$x{\in}\mathbb{Z}_{M}, y{\in}\mathbb{Z}_{N}$ and $\{|x{\rangle}\}$
is an orthonormal basis set for ${\cal H}_{M}$.  The sets
$\{|y{\rangle}\}$ and $\{|y{\oplus}f(x){\rangle}\}$ are, for any
fixed value of $f(x)$, orthonormal basis sets for ${\cal H}_{N}$.
These bases are the computational basis sets for both systems.  The
standard oracle operators may then be written as
\begin{equation}
\label{uf} U_{f}=\sum_{x{\in}\mathbb{Z}_{M},
y{\in}\mathbb{Z}_{N}}|x{\rangle}{\langle}x|{\otimes}|y{\oplus}f(x){\rangle}{\langle}y|.
\end{equation}
There are $N^{M}$ functions in ${\cal F}_{MN}$, so there are $N^{M}$
associated standard oracle operators $U_{f}$.  As we indicated
earlier, for any fixed $M,N$, these operators form an Abelian group.
To prove this, we observe that for two functions $f,f'{\in}{\cal
F}_{MN}$,
\begin{equation}
\label{uu} U_{f}U_{f'}=U_{f{\oplus}f'}.
\end{equation}
The standard oracle operator corresponding to the function
$\mathbb{Z}_{M}{\mapsto}0$ is the identity operator.  The inverse of
each standard oracle operator is also a standard oracle operator
because
\begin{equation}
\label{udag} U^{\dagger}_{f}=U_{0{\ominus}f},
\end{equation}
where ${\ominus}$ denotes subtraction modulo $N$.  These
observations, together with the associativity of modular addition,
prove the group property. The fact that this group is Abelian
follows from the simple observation that $f{\oplus}f'=f'{\oplus}f$,
i.e. modular addition is commutative. The commutativity of these
operators implies, as a consequence of Theorem \ref{commute}, that
there is no advantage to be gained by entangling the two systems
upon which the oracle operators act with other systems. Having said
that, there may be some advantage to be gained by entangling these
two systems with each other.

The standard oracle operators commute. They can therefore be
simultaneously diagonalised.  To do this, let us begin with the
$N$-dimensional Pegg-Barnett phase states \cite{PB}
\begin{equation}
|{\phi}_{Nn}{\rangle}=\frac{1}{\sqrt{N}}\sum_{y{\in}\mathbb{Z}_{N}}e^{\frac{2{\pi}iny}{N}}|y{\rangle}.
\end{equation}
The $N$-dimensional Pegg-Barnett phase operator (with zero reference
phase) is
\begin{equation}
{\Phi}_{N}=\sum_{n{\in}\mathbb{Z}_{N}}\frac{2{\pi}n}{N}|{\phi}_{Nn}{\rangle}{\langle}{\phi}_{Nn}|.
\end{equation}
The phase states $|{\phi}_{Nn}{\rangle}$ are, for each $N$,
orthonormal. They are the eigenstates of ${\Phi}_{N}$, having
corresponding eigenvalues $2{\pi}n/N$. Consider now the number shift
operator $e^{-i{\Phi}_{N}}$. This has the property
\begin{equation}
e^{-i{\Phi}_{N}}|y{\rangle}=|y{\oplus}1{\rangle}.
\end{equation}
Hence, we may write the standard oracle operator $U_{f}$ as
\begin{eqnarray}
\label{pboracle}
U_{f}&=&\sum_{x{\in}\mathbb{Z}_{M},
y{\in}\mathbb{Z}_{N}}|x{\rangle}{\langle}x|{\otimes}e^{-if(x){\Phi}_{N}}|y{\rangle}{\langle}y|
\nonumber
\\
&=&\sum_{x{\in}\mathbb{Z}_{M}}|x{\rangle}{\langle}x|{\otimes}e^{-if(x){\Phi}_{N}}\sum_{y{\in}\mathbb{Z}_{N}}|y{\rangle}{\langle}y| \nonumber \\
&=&\sum_{x{\in}\mathbb{Z}_{M}}|x{\rangle}{\langle}x|{\otimes}e^{-if(x){\Phi}_{N}}.
\end{eqnarray}
So we see that the state $|x{\rangle}{\otimes}|{\phi}_{Nn}{\rangle}$
is an eigenstate of $U_{f}$ with eigenvalue
$e^{\frac{-2{\pi}inf(x)}{N}}$.

\subsection{Minimal oracle operators}

Here we restrict our attention to invertible functions. We also
assume that $M=N$, in which case the functions will be permutations.
We can then consider simplified oracle operators of the form
\begin{equation}
\label{ro} Q_{f}|x{\rangle}=|f(x){\rangle}.
\end{equation}
Kashefi et al. \cite{KKVB} call these minimal oracle operators. They
are also known as erasing oracle operators in view of the fact that
they replace $x$ with $f(x)$.  Notice the connection here between
invertible functions and the invertibility of unitary operators.  We
may write these operators as
\begin{equation}
Q_{f}=\sum_{x{\in}\mathbb{Z}_{M}}|f(x){\rangle}{\langle}x|.
\end{equation}
Minimal oracle operators, unlike standard oracle operators, do not
generally commute with each other.  In fact, it is easy to show that
two minimal oracle operators commute iff the corresponding
permutations commute. It follows that Theorem \ref{commute} does not
apply to sets of such operators in general and we cannot rule out
the possibility that optimal discrimination among them may sometimes
require an entangled state. As such, it is appropriate to define the
entanglement-assisted minimal oracle operators
\begin{equation}
\bar{Q}_{f}=Q_{f}{\otimes}{\id}_{M}.
\end{equation}
Non-commuting minimal oracle operators are not simultaneously
diagonalisable.  Nevertheless, it is possible to diagonalise these
operators in general \cite{Atici}.  It is found that the eigenvalues
and eigenvectors depend on the cycle structure of the permutation.

As a consequence of the fact that minimal oracle operators only
exist for certain kinds of functions in ${\cal F}_{MN}$, whenever we
use the term oracle operators in this article, without specifying
whether or not they are standard, minimal, or entanglement-assisted
minimal oracle operators, unless it is clear from the context that
we are referring to all of these, it should be assumed that we are
referring to standard oracle operators \cite{Note1}.

\section{Condition for unambiguous discrimination among oracle operators}
\renewcommand{\theequation}{4.\arabic{equation}}
\setcounter{equation}{0}
\label{sec:4}
 From Section \ref{sec:2}, it is clear that we can obtain a necessary and sufficient condition for the unambiguous distinguishability of a set of either standard
 or minimal oracle operators if we can calculate the elements of
 the corresponding Gram matrix.  With this in mind, we can prove the
 following:

\begin{theorem}
\label{usdoracle} Consider a subset ${\sigma}{\subset}{\cal F}_{MN}$
with cardinality $K({\sigma})$.  We denote the functions in this set
by $f_{j}$, where $j=0,{\ldots},K({\sigma})-1$.  The standard and,
for permutations,  minimal oracle operators are denoted by
$U_{f_{j}}$ and $Q_{f_{j}}$ respectively. A necessary and sufficient
condition for the unambiguous distinguishability of either the
$U_{f_{j}}$ or the $Q_{f_{j}}$ is
\begin{equation}
\label{detgamma}
\det({\Gamma})>0
\end{equation}
where ${\Gamma}=({\Gamma}_{j'j})$ is a
$K({\sigma}){\times}K({\sigma})$ matrix with elements
\begin{equation}
\label{Gamma}
{\Gamma}_{j'j}=\sum_{x{\in}\mathbb{Z}_{M}}{\langle}f_{j'}(x)|f_{j}(x){\rangle}.
\end{equation}
\end{theorem}
\noindent{\bf Proof}:  Beginning with standard oracle operators, Eq.
\eqref{uf} implies that the elements of the Gram matrix of these
operators are
\begin{eqnarray}
\mathrm{Tr}(U^{\dagger}_{f_{j'}}U_{f_{j}})&=&\mathrm{Tr}\left(\sum_{x,x'{\in}\mathbb{Z}_{M},
y,y'{\in}\mathbb{Z}_{N}}\left(|x'{\rangle}{\langle}x'|{\otimes}|y'{\rangle}{\langle}y'{\oplus}f_{j'}(x')|\right)\left(|x{\rangle}{\langle}x|{\otimes}|y{\oplus}f_{j}(x){\rangle}{\langle}y|\right)\right)
\nonumber \\ &=&\mathrm{Tr}\left(\sum_{x{\in}\mathbb{Z}_{M},
y,y'{\in}\mathbb{Z}_{N}}{\langle}y'{\oplus}f_{j'}(x)|y{\oplus}f_{j}(x){\rangle}|x{\rangle}{\langle}x|{\otimes}|y'{\rangle}{\langle}y|\right)
\nonumber \\ &=&\mathrm{Tr}\left(\sum_{x{\in}\mathbb{Z}_{M},
y,y'{\in}\mathbb{Z}_{N}}{\langle}f_{j'}(x)|e^{-i{\Phi}_{N}(y-y')}|f_{j}(x){\rangle}|x{\rangle}{\langle}x|{\otimes}|y'{\rangle}{\langle}y|\right)
\nonumber \\
&=&N\sum_{x{\in}\mathbb{Z}_{M}}{\langle}f_{j'}(x)|f_{j}(x){\rangle}=N{\Gamma}_{j'j},
\end{eqnarray}
where we have made use of Eq. \eqref{pboracle}.  From this, we see
that the matrix ${\Gamma}$ defined in Eq. \eqref{Gamma} is
proportional to the Gram matrix defined in Eq. \eqref{Grammatrix}.
As such, these two matrices will be non-singular under the same
circumstances.

Let us now turn to the minimal oracle operators. Anticipating the
discussions of later sections, it will be more convenient to work
with the entanglement-assisted minimal oracle operators
$\bar{Q}_{f_{j}}$ instead.  The Gram matrix of these operators is
proportional to that of the unassisted minimal oracle operators
$Q_{f_{j}}$, so they are non-singular under the same conditions. We
find that
\begin{eqnarray}
\mathrm{Tr}(\bar{Q}^{\dagger}_{f_{j'}}\bar{Q}_{f_{j}})&=&\mathrm{Tr}(Q^{\dagger}_{f_{j'}}Q_{f_{j}}{\otimes}{\id}_{M})
\nonumber \\
&=&M\mathrm{Tr}\left(\sum_{x,x'{\in}\mathbb{Z}_{M}}{\langle}f_{j'}(x')|f_{j}(x){\rangle}|x'{\rangle}{\langle}x|\right)
\nonumber \\
&=&M\sum_{x{\in}\mathbb{Z}_{M}}{\langle}f_{j'}(x)|f_{j}(x){\rangle}=M{\Gamma}_{j'j}.
\end{eqnarray}
Making the identification $M=N$, we see that this is exactly the
result we obtained for the standard oracle operators.  So, Eq.
\eqref{detgamma} is a necessary and sufficient condition for the
unambiguous distinguishability of the standard and the minimal
oracle operators. This completes the
proof. ${\Box}$\\

Evidently, the Gram matrices of the standard and
entanglement-assisted minimal oracle operators are not merely
proportional to each other. They are in fact identical.  We shall
explore the implications of this in Section \ref{sec:8}.

Given its significance, it would be desirable to have a suitably
transparent interpretation of the matrix ${\Gamma}$. Fortunately,
such an interpretation is possible.  From its definition in Eq.
\eqref{Gamma}, it is readily apparent that ${\Gamma}_{j'j}$ is equal
to the number of values of $x$ for which $f_{j'}(x)=f_{j}(x)$.  As
such, the magnitude of each element of ${\Gamma}$ quantifies the
indistinguishability of the corresponding pair of functions. It
follows from this observation that our condition for unambiguous
oracle operator discrimination does not depend on any specifically
quantum mechanical properties of the oracle operators. It can be
understood solely in terms of pairwise relationships between the
functions that these operators compute.

One final point to note is that all elements of ${\sigma}$ are
distinct because all elements of ${\cal F}_{MN}$ are distinct.
Throughout this paper, we only consider sets of functions that are a
priori distinct from each other (i.e. we do not consider functions
that are identical and merely given different labels.)  For
identical functions, the corresponding oracle operators would also
be, in a given model, identical and therefore uninteresting in terms
of their distinguishability properties.

\section{Some consequences of the unambiguous oracle operator discrimination condition}
\renewcommand{\theequation}{5.\arabic{equation}}
\setcounter{equation}{0} \label{sec:5}
\subsection{Classical discrimination among functions and unambiguous oracle operator discrimination}

Here, we shall describe some interesting consequences of the
unambiguous oracle operator discrimination condition derived in
Section \ref{sec:4}.  In order to place quantum oracle operator
discrimination in context, it is important to understand the
conditions under which discrimination among the associated functions
can be achieved classically. Indeed, this issue will become
even more important in Section \ref{sec:6}.  In what follows, we will make use of the following definition:\\

\begin{definition}[Classical Distinguishability]
Consider a subset ${\sigma}{\subset}{\cal F}_{MN}$.  We say that the
functions $f_{j}{\in}{\sigma}$ are classically distinguishable iff
there exists $x_{0}{\in}\mathbb{Z}_{M}$ such that
\begin{equation}
f_{j'}(x_{0}){\neq}f_{j}(x_{0})\;\;\mathrm{if}\;\;j'{\neq}j.
\end{equation}
\end{definition}
This definition formalises the intuitive notion that for the
functions $f_{j}$ to be distinguishable if computed classically,
there must be at least one value of the input variable $x$, which we
have denoted by $x_{0}$, for which the $f_{j}$ all have different
values.\\

As an application of this definition, consider functions computed
using the following reversible classical oracle:
\begin{equation}
\label{classoracle} (x,y)\mapsto(x,y{\oplus}f(x)).
\end{equation}
Here, $x,y$ and $f(x)$ are assumed to be known.  This oracle is the
natural classical equivalent of the standard quantum oracle operator
for $f(x)$. That the distinguishability of functions computed using
this oracle accords with the above definition is obvious if $y=0$.
For $y{\neq}0$, it can be seen as a simple consequence of the
invertibility of modular addition.

It is also clear that a standard quantum oracle operator acting on
the computational basis state $|x{\rangle}{\otimes}|y{\rangle}$ will
result in another computational basis state whose labels are
transformed according to Eq. \eqref{classoracle}. It follows that if
a set of functions are classically distinguishable, then their
standard oracle operators are perfectly distinguishable. They are
therefore, obviously, unambiguously distinguishable.

This observation is straightforward.  However, it is nevertheless
interesting to see how Theorem \ref{usdoracle} can be used to
directly show that classical distinguishability implies unambiguous
distinguishability of the corresponding oracle operators, for the
purpose of helping us become acquainted with the ways in which this
condition can be used.

To do so, consider the positive semi-definite matrix
${\Gamma}_{x}=({\langle}f_{j'}(x)|f_{j}(x){\rangle})$.  Clearly,
${\Gamma}=\sum_{x{\in}\mathbb{Z}_{M}}{\Gamma}_{x}$.  If the
functions are classically distinguishable for some $x=x_{0}$, then
${\Gamma}_{x_{0}}={\id}_{M}$, which is obviously non-singular. The
positive semi-definiteness of the ${\Gamma}_{x}$ implies that if any
of them are non-singular, then ${\Gamma}$ is non-singular also. This
shows how classical distinguishability confirms unambiguous
distinguishability of the corresponding oracle operators.

\subsection{Limitations on unambiguous
discrimination among all standard oracle operators for fixed $M,N$}

One interesting question concerns unambiguous discrimination among
the standard oracle operators corresponding to all functions in
${\cal F}_{MN}$, for arbitrary, fixed integers $M,N{\geq}1$.  We
would like to know whether or not this can be achieved.
Unfortunately, except in certain trivial cases, this is not
possible.  We can prove:
\begin{theorem}
The standard oracle operators corresponding to all functions in
${\cal F}_{MN}$ are not unambiguously distinguishable for any fixed
$M$ and $N$ both ${\geq}2$.
\end{theorem}

\noindent{\bf Proof}:  We treat the case $M=N=2$ first. Here, we
have the following four functions:
\begin{eqnarray}
\label{f221}
f_{0}:(0,1)&{\mapsto}&(0,0), \\
\label{f222}
f_{1}:(0,1)&{\mapsto}&(0,1), \\
\label{f223}
f_{2}:(0,1)&{\mapsto}&(1,0), \\
\label{f224}
f_{3}:(0,1)&{\mapsto}&(1,1).
\end{eqnarray}
From this, we easily obtain
\begin{equation}
\label{f22gamma}
{\Gamma}= \left( {\begin{array}{cccc}
2 & 1 & 1 & 0 \\
1 & 2 & 0 & 1 \\
1 & 0 & 2 & 1 \\
0 & 1 & 1 & 2
\end{array}}
 \right).
\end{equation}
One can confirm that this matrix is singular, e.g. by direct
calculation of its determinant. This implies that the standard
oracle operators corresponding to the four functions in ${\cal
F}_{22}$ are not unambiguously distinguishable.

For higher values of $M$ and $N$, we make use of inequality
\eqref{kd}. When the set of possible functions is ${\cal F}_{MN}$ we
have ${\sigma}={\cal F}_{MN}$ and therefore $K({\sigma})=N^{M}$. The
dimensionality of the Hilbert space upon which the standard oracle
operators act is $MN$.  It follows that if these operators are to be
unambiguously distinguishable, we must have
\begin{equation}
\label{mnineq}
M{\geq}N^{M-1}.
\end{equation}
It is a straightforward matter to show that this inequality cannot
be satisfied for $M{\geq}2$ and $N>2$ or $N{\geq}2$ and $M>2$.  To
do so, consider the function
\begin{equation}
g(M,N)=N^{M-1}-M.
\end{equation}
We note firstly that $g(2,2)=0$.  We now show that for $M{\geq}2$
and $N{\geq}2$, $g(M,N)$ is increasing with respect to both of these
variables.  To do so, we observe that
\begin{equation}
\frac{{\partial}g(M,N)}{{\partial}M}=(M-1)N^{M-2},
\end{equation}
which is strictly positive for $M{\geq}2$ and all positive $N$.  We
also see that
\begin{equation}
\frac{{\partial}g(M,N)}{{\partial}N}=N^{M-1}{\ln}(N)-1
{\geq}2{\ln}(2)-1>0,
\end{equation}
for $M{\geq}2$ and $N{\geq}2$, proving our assertion. $\Box$\\

The remaining situations to consider are when either or both
$M,N=1$. If $M=1$, then it is easily seen that all functions in
${\cal F}_{1N}$ are distinguishable in order to be distinct, which
they are. This situation is somewhat trivial.  If, on the other
hand, $N=1$ then there is only one possible function, which is
obviously known and so this situation is also trivial.  We then see
that except in these trivial cases, the set of all corresponding
standard oracle operators for fixed $M,N$ cannot be unambiguously
discriminated.

\subsection{Unambiguous discrimination among Grover
oracle operators}

We have seen that the standard oracle operators corresponding to
classically distinguishable functions are perfectly and therefore
unambiguously distinguishable.  Naturally, one would like to know
whether or not the converse is true, that is, if the standard oracle
operators for a set of functions are unambiguously distinguishable,
then are the functions classically distinguishable?  That this is
not generally the case can be concluded on the basis of the fact
that the standard oracle operators corresponding to one of the most
important quantum algorithms, the Bernstein-Vazirani algorithm, are
perfectly distinguishable while the corresponding functions are not,
in general, classically distinguishable \cite{bv}.

We will show here that the standard oracle operators which arise
through the consideration of another important quantum algorithm,
namely Grover's famous quantum search algorithm \cite{Grover}, are
unambiguously distinguishable even though the corresponding
functions are not classically distinguishable.  For an unsorted
database with $M$ items, these functions, which are elements of
${\cal F}_{M2}$, are
\begin{equation}
\label{groverfunctions}
f_{j}(x)={\delta}_{xj},
\end{equation}
where $j=0,{\ldots},M-1$.  For $M{\geq}3$, these functions are not
classically distinguishable.  We have $K({\sigma})=M$ functions in
this set. These functions are such that $f_{j}(x)=0$ for all $x$
except $x=j$, in which case $f_{j}(x)=1$.

For the corresponding standard oracle operators to be linearly
independent and thus unambiguously distinguishable, we require that
${\Gamma}$ is non-singular.  For general $M$, the elements of this
matrix are easily computed and we find that
\begin{equation}
{\Gamma}_{j'j}=2{\delta}_{j'j}+(M-2)
\end{equation}
and that the matrix itself may be written as
\begin{equation}
{\Gamma}=2{\id}_{M}+M(M-2)P[{\chi}],
\end{equation}
where ${\id}_{M}$ is the identity matrix on $\mathbb{C}^{M}$ and
$P[{\chi}]$ is the projector onto the subspace spanned by the
$M$-component, normalised vector ${\chi}=M^{-1/2}(1,1,..,1)$.  This
matrix therefore has two distinct eigenvalues: 2, which is
$(M-1)$-fold degenerate, and $(M-1)^{2}+1$. These are all non-zero
and so ${\Gamma}$ is non-singular.  Indeed the determinant of
${\Gamma}$, being their product, is
\begin{equation}
\det({\Gamma})=2^{M-1}[(M-1)^{2}+1]>0,
\end{equation}
implying that for all $M$, the standard Grover oracle operators are
unambiguously distinguishable.

It is interesting to consider the unambiguous distinguishability of
the oracle operators used by Grover in his original exposition of
his algorithm.  These may be seen to emerge from the corresponding
standard oracle operators in the following way.  We have $N=2$ so
let us consider preparing the second system upon which the standard
oracle operators act in the state
$|-{\rangle}=\frac{1}{\sqrt{2}}(|0{\rangle}-|1{\rangle})$.  One can
then verify that
\begin{equation}
U_{f_{j}}|x{\rangle}{\otimes}|-{\rangle}=G_{j}|x{\rangle}{\otimes}|-{\rangle},
\end{equation}
where $G_{j}$ is the original Grover oracle operator whose action
can be described in the following way:
\begin{equation}
G_{j}|x{\rangle}=(-1)^{{\delta}_{xj}}|x{\rangle}.
\end{equation}
The operator $G_{j}$ clearly imparts a ${\pi}$ phase shift to the
state corresponding to the sought for item and leaves the states
corresponding to other items invariant.  Actually, it is interesting
to generalise this to an arbitrary phase shift in the manner of Long
et al.\cite{Long1}.  These authors considered the operators
$G_{j}({\theta})$, which act in the following way:
\begin{equation}
G_{j}({\theta})|x{\rangle}=\left[{\delta}_{jx}(e^{i{\theta}}-1)+1\right]|x{\rangle}
\end{equation}
which impart an arbitrary phase shift ${\theta}$ instead.  One can
readily verify that $G_{j}({\pi})=G_{j}$.  The unambiguous
distinguishability of these operators is determined by the
determinant of their Gram matrix $G=(G_{j'j})$.  We find that the
Gram matrix elements $G_{j'j}$ are
\begin{equation}
G_{j'j}=\mathrm{Tr}(G^{\dagger}_{j'}G_{j})=M+2(1-{\delta}_{j'j})({\cos}({\theta})-1)
\end{equation}
and thus the Gram matrix $G$ may be written as
\begin{equation}
G=2(1-{\cos}({\theta})){\id}_{M}+M(M-2+2{\cos}({\theta}))P[{\chi}].
\end{equation}
The eigenvalues of $G$ are then $2(1-{\cos}({\theta}))$, which is
$(M-1)$-fold degenerate and $M^{2}+2(1-M)(1-{\cos}({\theta}))$,
which is non-degenerate. The determinant of $G$ is then seen to be
\begin{equation}
{\det}(G)=2^{M-1}(1-{\cos}({\theta}))^{M-1}[(M-1+{\cos}({\theta}))^{2}+{\sin}^{2}({\theta})]
\end{equation}
which is non-zero for all values of ${\theta}$ which are not integer
multiples of $2{\pi}$.  So, we see that the oracle operators
$G_{j}({\theta})$ are unambiguously distinguishable for any
${\theta}{\neq}2k{\pi}$, $k{\in}\mathbb{Z}$.

We have seen, in terms of both the standard and original Grover
oracle operators, that it is possible to unambiguously find an
unknown marked item in an arbitrarily large unsorted database with
one query.  This contrasts strongly with the situation that arises
if we require the search to be carried out deterministically. It was
shown by Boyer et al. \cite{BBHT} that a one-query deterministic
Grover-type search of an unsorted database is only possible if there
are ${\leq}4$ items.

\section{Unambiguous oracle operator discrimination for totally indistinguishable
functions}
\renewcommand{\theequation}{6.\arabic{equation}}
\renewcommand{\thetable}{\arabic{table}}
\setcounter{equation}{0} \setcounter{table}{0} \label{sec:6}

\subsection{General considerations}

We saw in the preceding section that the Grover oracle operators are
unambiguously distinguishable for an arbitrarily large database. One
point worth making is that a limited form of unambiguous, indeed
perfect, distinguishability holds for the analogous classical
situation.  If we perform a classical search of a large, unsorted
database, then even with one shot, there will be a finite
probability of obtaining the desired item.  In terms of the
functions $f_{j}(x)$ in Eq. \eqref{groverfunctions}, this is
equivalent to saying that, for any $x$ and with a suitable initial
state that depends on $x$, when we evaluate the unknown $f_{j}(x)$,
there is a non-zero probability that the result of this function
evaluation will be 1. When this is so, we can uniquely identify
which function was computed, since there is a one-to-one
correspondence between the functions $f_{j}(x)$, or more precisely
the value of the index $j$, and the value of $x$ for which
$f_{j}(x)=1$. Of course, classically, for an unsorted database of
$>2$ items, when we obtain a value of $0$, we cannot determine which
function was computed.  In this scenario, for each choice of $x$,
there is only one function that can be conclusively identified.

The strength of unambiguous standard oracle discrimination in
relation to this possibility is that a conclusive discrimination
among all oracle operators/functions is possible with one shot and a
fixed input state. Nevertheless, the fact that the above scenario is
possible leads to the following question: are there functions among
which we can never discriminate classically, yet for which the
corresponding standard oracle operators are unambiguously
distinguishable?

To be precise about what we mean by functions among which we can
never discriminate classically, we shall employ the following:
\begin{definition}[Totally indistinguishable functions]
Consider a set of functions ${\sigma}{\subset}{\cal F}_{MN}$.  This
set is totally indistinguishable iff, for any $x{\in}\mathbb{Z}_{M}$
and for any function $f_{j}{\in}{\sigma}$, there a function
$f_{j'}{\in}{\sigma}$, where $j'{\neq}j$, such that
$f_{j}(x)=f_{j'}(x)$.
\end{definition}
Informally, a totally indistinguishable set of functions is a set
such that, for any value of the input variable $x$, there will be at
least two functions which produce the same output.  So, with a
knowledge of only $x$ and the value of the function computed for
this value, we can never determine which function was computed.

One elementary observation we can make about totally
indistinguishable functions is

\begin{lemma}
\label{lem:four} Let ${\sigma}{\subset}{\cal F}_{MN}$, having
cardinality $K({\sigma})$. If this set is totally indistinguishable
then $K({\sigma}){\geq}4$.
\end{lemma}
\noindent{\bf Proof}: All functions in ${\sigma}$ are distinct. If
two functions $f_{0}(x)$ and $f_{1}(x)$ are distinct, then there
exists $x$ such that $f_{0}(x){\neq}f_{1}(x)$. For such a value of
$x$, $f_{0}$ and $f_{1}$ are clearly classically distinguishable.
This shows that two distinct functions cannot be totally
indistinguishable.

In the case of three functions  $f_{0},f_{1}$ and $f_{2}$, for these
functions to be distinct, there must exist $x$ such that
$f_{0}(x),f_{1}(x)$ and $f_{2}(x)$ are not all equal.  It is,
however, impossible to have three numbers which are not all equal
yet where each one is equal to one of the other two, which would be
required for the functions to be totally indistinguishable. This
proves that there are no sets of two or three distinct functions
which possess
total indistinguishability and that such sets must therefore consist of at least four functions.$\Box$\\

\subsection{The case $M=2$}
We have seen that for a set of functions to possess total
indistinguishability, there must be at least four functions in this
set. Here we shall prove a constraint on $M$, the number of possible
values of the independent variable $x$, which limits the conditions
under which the corresponding standard oracle operators can be
unambiguously discriminated. Clearly, the lowest, non-trivial value
that $M$ may assume is 2, so let us investigate functions in ${\cal
F}_{2N}$. These are functions of the form
\begin{equation}
\label{ab} f:(0,1){\mapsto}(a,b),
\end{equation}
where $a,b{\in}\mathbb{Z}_{N}$ for arbitrary integer $N{\geq}2$.

One interesting property of the elements of ${\cal F}_{2N}$, which
is readily appreciated from this expression, is that they can be
represented as points in a two-dimensional plane.  More
specifically, let us consider $\mathbb{R}^{2}$ endowed with a
Cartesian coordinate system $(X,Y)$.  We capitalise these
coordinates as $x$ and $y$ are already in use.  Let us now imagine
an $N{\times}N$ lattice of points in the first quadrant (including
the origin and coordinate axes.) These points are at locations where
both coordinates take integer values. One can easily see that there
is a one-to-one correspondence between these points and the elements
of ${\cal F}_{2N}$.  For $f$ given by Eq. \eqref{ab}, its
corresponding point has coordinates $(X,Y)=(a,b)$.

This representation provides an appealing of way of visualising the
property of total indistinguishability.  We know that a set of
functions ${\sigma}{\subset}{\cal F}_{2N}$ of this nature must have
the property that, for any $f_{j}{\in}{\sigma}$, there exist
different functions $f_{k},f_{l}{\in}{\sigma}$ such that
$f_{j}(0)=f_{k}(0)$ and $f_{j}(1)=f_{l}(1)$.  This translates in our
geometric representation to the requirement that, in our set of
points, there is no line parallel to either the $X$- or $Y$-axis
which is occupied by only one of these points.  In each occupied
line there must be at least two of them.

We are led by this observation to the following graphical
representation of the functions in ${\sigma}$.  Let us, with a
slight abuse of notation, define an undirected graph $G({\sigma})$
in the $(X,Y)$-plane, whose vertices are the points we have been
describing. Vertex $V_{j}$ corresponds to the function $f_{j}$.
Edges occur between vertices with either the same $X$- or
$Y$-coordinate, that is, if the corresponding functions give the
same value when evaluated on either $0$ or $1$. We shall say that
vertices with the same $X$ or $Y$ coordinate are $Y$- or
$X$-adjacent respectively, because for two $X$-adjacent points, the
edge will run in the $Y$ direction and vice versa.  They are
adjacent if they are either $X$- or $Y$-adjacent.  Clearly, no two
distinct points can be both $X$- and $Y$-adjacent.  For a set of
totally indistinguishable functions ${\sigma}$, we can deduce
that the graph $G({\sigma})$ has the following properties: \\

\noindent(i) Each vertex in $G({\sigma})$ has degree ${\geq}2$. \\

\noindent(ii) Each connected component \cite{ccomp} of $G({\sigma})$
corresponds to a subset of ${\sigma}$ which is
totally indistinguishable.  \\

\noindent(iii) As a consequence of Lemma \ref{lem:four}, each
connected component of
$G({\sigma})$ has at least four vertices. \\

\noindent(iv) If vertices $V_{j}$ and $V_{k}$ are $X$-adjacent and $V_{j}$ and $V_{l}$ are $Y$-adjacent, then $V_{k}$ and $V_{l}$ are not adjacent. \\

\noindent(v) The adjacency matrix $A$ of the graph $G({\sigma})$ and
the matrix ${\Gamma}$ obtained from the corresponding
functions/standard oracle operators are related through
\begin{equation}
\label{gammaA}
{\Gamma}=A+2{\id}_{K({\sigma})},
\end{equation}
where ${\id}_{K({\sigma})}$ is the $K({\sigma}){\times}K({\sigma})$
identity
matrix. \\

\noindent(vi)  More generally, consider an arbitrary subset
${\sigma}'{\subset}{\sigma}$ with complement $\bar{\sigma}'$ within
${\sigma}$.  The graph $G({\sigma}')$ obtained from $G({\sigma})$ by
deleting all vertices corresponding to functions in $\bar{\sigma}'$
and all edges connecting these to vertices corresponding to
functions in ${\sigma}'$ is an induced subgraph \cite{ccomp} of
$G({\sigma})$. Any induced subgraph of $G({\sigma}')$ can be
constructed in this manner. A matrix $\tilde{\Gamma}$ is constructed
for the functions in ${\sigma}'$ in the same way as ${\Gamma}$ is
constructed for those in ${\sigma}$ in Eq. \eqref{Gamma}.  The
adjacency matrix of $G({\sigma}')$, which we shall denote by
$\tilde{A}$, is related to $\tilde{\Gamma}$ through
\begin{equation}
\label{tildegammaa} \tilde{\Gamma}=\tilde{A}+2{\id}_{K({\sigma}')},
\end{equation}
where ${\id}_{K({\sigma}')}$ is the
$K({\sigma}'){\times}K({\sigma}')$ identity
matrix.  Moreover, $\tilde{\Gamma}$ is a principal submatrix of ${\Gamma}$, implying that if $\tilde{\Gamma}$ is singular, then ${\Gamma}$ is singular also.\\

\begin{figure}
\begin{center}
\epsfxsize8cm \centerline{\epsfbox{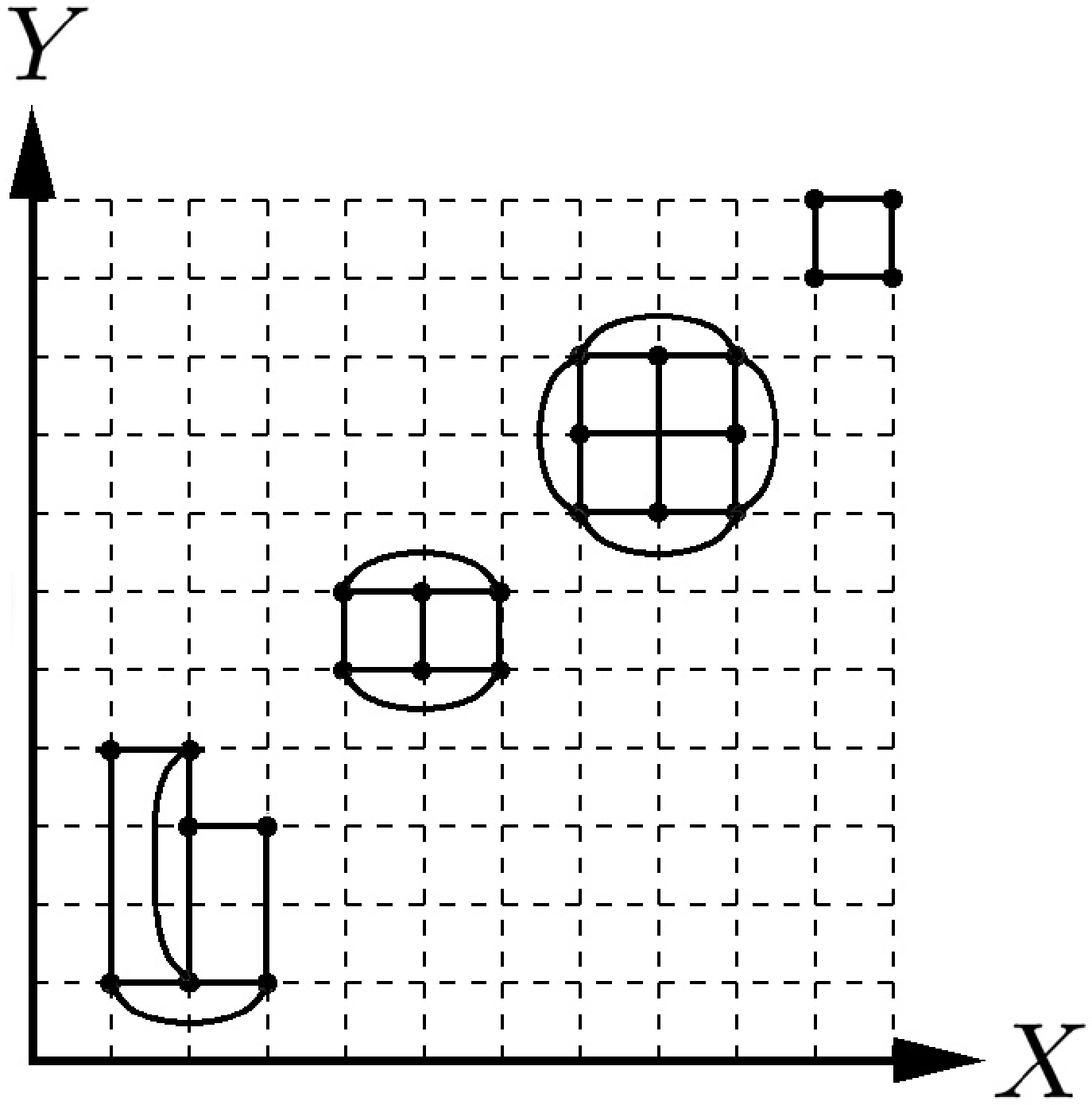}}
\end{center}
\caption{Example of a graph $G({\sigma})$ corresponding to a finite
set ${\sigma}$ of totally indistinguishable functions in ${\cal
F}_{2N}$.  This example illustrates the adjacency and connectivity
phenomena which characterise such graphs in general.}
\label{figure1}
\end{figure}

Figure \ref{figure1} depicts a typical $G({\sigma})$ corresponding
to a totally indistinguishable set ${\sigma}$, illustrating features
which are typical of such graphs.  Having established this graphical
framework, we are now in a position to use it to prove our main
result for functions in ${\cal F}_{2N}$:

\begin{theorem}
\label{twotif} Let ${\sigma}{\subset}{\cal F}_{2N}$ be a finite set
of totally indistinguishable functions. Then the standard oracle
operators corresponding to them are not unambiguously
distinguishable.
\end{theorem}

\noindent{\bf Proof}: Our approach to proving this is as follows. We
know that the standard oracle operators will not be unambiguously
distinguishable iff ${\Gamma}$ is singular, i.e. if one of its
eigenvalues is zero.  From Eq. \eqref{gammaA}, we see that this is
equivalent to the adjacency matrix $A$ having eigenvalue $-2$. It is
impractical to determine the universal existence of this eigenvalue
by attempting to diagonalise the adjacency matrices of all graphs
corresponding to sets of totally indistinguishable functions in
${\cal F}_{2N}$, for finite $N$. Instead, we will make use of
property (vi) above.   This property, in particular Eq.
\eqref{tildegammaa}, implies that if there exists an induced
subgraph $G({\sigma}')$ of $G({\sigma})$ whose adjacency matrix
$\tilde{A}$ has eigenvalue $-2$, then the matrix $\tilde{\Gamma}$ is
singular. This will in turn imply that ${\Gamma}$ is singular.

One eminently simple class of graphs whose adjacency matrices have
eigenvalue $-2$ are even circulant, or cycle graphs.  A cycle graph
with $K$ vertices is a graph consisting of a single cycle linking
all of its vertices. These vertices can be labeled in such a way
that the adjacency structure is simply that vertex $V_{j+1}$ is
adjacent to vertex $V_{j}$ and that vertex $V_{K-1}$ is adjacent to
$V_{0}$. Moreover, such a vertex relabeling corresponds to a
similarity transformation of the adjacency matrix $\tilde{A}$ by an
orthogonal matrix $O$, which leaves its spectrum invariant.  It
follows that for such a cycle graph with adjacency matrix
$\tilde{A}$, we have

\begin{equation}
O\tilde{A}O^{T}=\begin{pmatrix}
0 & 1 & & & 1 \\
1 & 0 & {\ddots} & & & \\
& {\ddots} & {\ddots} & {\ddots} & & \\
&  & {\ddots} & 0 & 1 \\
1 & & & 1 & 0
\end{pmatrix},
\end{equation}
where $T$ denotes transposition and the entries not specified are
zero. This matrix is a circulant matrix.  Many properties of
circulant matrices are well-established, see e.g. \cite{Gray}.  In
particular, the eigenvalues of the above matrix are
\begin{equation}
{\lambda}_{r}=2{\cos}\left(\frac{2{\pi}r}{K}\right),\;\;\;\;r=0,{\ldots},K-1.
\end{equation}
One readily finds that for even $K$, we have ${\lambda}_{K/2}=-2$.
So, the adjacency matrix of every cycle graph with an even number of
vertices has $-2$ as one of its eigenvalues.

From this, we see that we will be able to complete the proof if we
can show that every graph $G({\sigma})$ corresponding to a finite
set of totally indistinguishable functions ${\sigma}{\subset}{\cal
F}_{2N}$ has an induced subgraph $G({\sigma}')$ which is an even
cycle. We are indeed able to show this.  Formally, we have

\begin{figure}
\begin{center}
\epsfxsize8cm \centerline{\epsfbox{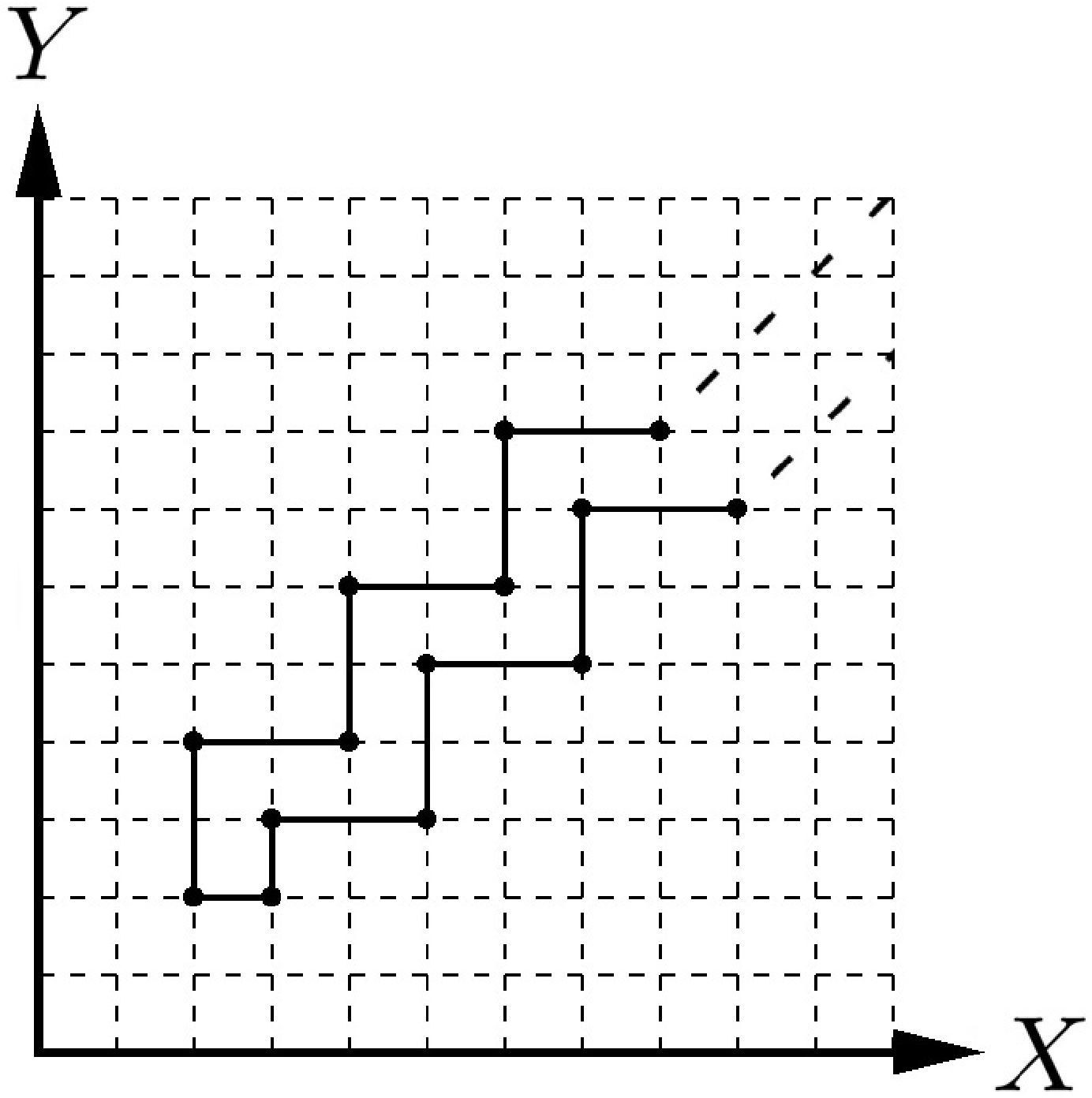}}
\end{center}
\caption{Example of an acyclic graph $G({\sigma})$ corresponding to
an infinite set of totally indistinguishable functions in ${\cal
F}_{2N}$, where $N=+{\infty}$.} \label{figure2}
\end{figure}

\begin{theorem}
\label{cycle} Every connected component of a graph $G({\sigma})$,
where ${\sigma}{\subset}{\cal F}_{2N}$ is a finite set of totally
indistinguishable functions, has an induced subgraph which is an
even cycle of length ${\geq}4$.
\end{theorem}

Our proof of this is somewhat intricate.  As such, we have placed it
in the Appendix.  Since every induced subgraph of a connected
component of $G({\sigma})$ is itself an induced subgraph of
$G({\sigma})$, we obtain the desired result, that the standard
oracle operators for any finite set of totally indistinguishable
functions
in ${\cal F}_{2N}$ are not unambiguously distinguishable. $\Box$\\

This result provides an intriguing demonstration of the global
implications of a local phenomenon.  The total indistinguishability
property is local, since it is a constraint on the adjacency
properties of $G({\sigma})$.  However, we have seen that this
implies the existence of global, indeed topological features, namely
induced even-length cycles.  To our knowledge, this is the first
demonstration of the relevance of topology to unambiguous operator
or state discrimination.

One further point worth noting is that our proof of Theorem
\ref{twotif} relies on Theorem \ref{cycle}, which in turn depends
upon the assumption that the set ${\sigma}$ is finite.  When this is
not the case, one is able to obtain a set of totally
indistinguishable functions ${\sigma}$ whose corresponding graph
$G({\sigma})$ is acyclic (contains no cycles.) A simple example of
such a graph is given in Figure \ref{figure2}.

\subsection{The case $K({\sigma})=4$}
It follows from the foregoing results that if we wish to obtain a
finite set of totally indistinguishable functions with unambiguously
distinguishable standard oracle operators, we require $M{\geq}3$ and
$K({\sigma}){\geq}4$.  Here, we shall see that such sets of
functions do indeed exist.  In fact, we will give a complete
characterisation of all such sets of four functions in ${\cal
F}_{MN}$, for all fixed $M{\geq}3$ and $N{\geq}1$.

To begin with, we write four arbitrary functions in ${\cal F}_{MN}$
as
\begin{eqnarray}
f_{0}:(0,1,{\ldots},M-1)&{\mapsto}&(a_{0},a_{1},{\ldots},a_{M-1}), \\
f_{1}:(0,1,{\ldots},M-1)&{\mapsto}&(b_{0},b_{1},{\ldots},b_{M-1}), \\
f_{2}:(0,1,{\ldots},M-1)&{\mapsto}&(c_{0},c_{1},{\ldots},c_{M-1}), \\
f_{3}:(0,1,{\ldots},M-1)&{\mapsto}&(d_{0},d_{1},{\ldots},d_{M-1}),
\end{eqnarray}
for some $a_{x},b_{x},c_{x},d_{x}{\in}\mathbb{Z}_{N}$ where
$x{\in}\mathbb{Z}_{M}$.  It will be convenient to treat the right
hand side of this expression as a $4{\times}M$ matrix which we shall
refer to as the function matrix.  We assume at the outset that these
functions are totally indistinguishable. Of particular importance to
us will be the columns of this matrix. The form that these columns
may take is strongly constrained by the total indistinguishability
condition. This implies that, in each column, all elements either
have the same value, or that there are two different values in each
column, with each occurring twice. From this, we find that for a
given set of four functions in ${\cal F}_{MN}$, each column can take
one of four possible forms. These are shown in Table 1.  We have
numbered the four column types $1,{\ldots},4$.  We have also denoted
by $N_{i}$ the number of occurrences of column type $i$ in the
function matrix, where $i=1,{\ldots},4$. Clearly, we have
\begin{equation}
\label{summi} \sum_{i=1}^{4}N_{i}=M.
\end{equation}
We point out that in the notation of the table, $\bar{a}_{x}$ is
some number in $\mathbb{Z}_{N}$ which is not equal to $a_{x}$.

\begin{table}[h]
\label{table1}
\begin{center}
\begin{tabular}{|c|c|c|c|}
\hline
$a_{x}$ & $a_{x}$ & $a_{x}$ & $a_{x}$ \\
$a_{x}$ & $a_{x}$ & $\bar{a}_{x}$ & $\bar{a}_{x}$ \\
$a_{x}$ & $\bar{a}_{x}$ & $a_{x}$ & $\bar{a}_{x}$ \\
$a_{x}$ & $\bar{a}_{x}$ & $\bar{a}_{x}$ & $a_{x}$ \\
\hline
$N_{1}$ & $N_{2}$ & $N_{3}$ & $N_{4}$ \\
\hline
\end{tabular}
\end{center}
\caption{Forms of the four possible column types in the function
matrix for a set of four totally indistinguishable functions.  The
$N_{i}$ are the frequencies of each of these column types in the
function matrix.}
\end{table}
The importance of the role played by these four column types is
illustrated by the fact that the elements of the matrix ${\Gamma}$
can be expressed solely in terms of their frequencies $N_{i}$.
Making use of Eq. \eqref{summi}, we find that
\begin{equation}
{\Gamma}= \left( {\begin{array}{cccc}
M & N_{1}+N_{2} & N_{1}+N_{3} & N_{1}+N_{4} \\
N_{1}+N_{2} & M & N_{1}+N_{4} & N_{1}+N_{3} \\
N_{1}+N_{3} & N_{1}+N_{4} & M & N_{1}+N_{2} \\
N_{1}+N_{4} & N_{1}+N_{3} & N_{1}+N_{2} & M
\end{array}}
 \right).
\end{equation}
The determinant of this matrix is readily evaluated.  Again with the
aid of Eq. \eqref{summi}, we obtain
\begin{equation}
{\det}({\Gamma})=16(M+N_{1})N_{2}N_{3}N_{4}.
\end{equation}
From this, we see that ${\Gamma}$ will be non-singular iff
\begin{equation}
\label{niineq} N_{2},N_{3},N_{4}>0.
\end{equation}
This is a completely general condition, expressed in terms of the
frequencies of the column types in the function matrix, for four
totally indistinguishable functions to have unambiguously
distinguishable standard oracle operators.  It states that all three
column types corresponding to the four functions not all having
equal values must occur.

This condition has the appealing and intuitively expected property
of being symmetrical with respect to these three column types.  This
property can be understood as arising from the fact that, if we
relabel the last three columns in Table 1 amongst themselves, then
this can be seen to be equivalent to permuting the labels of the
functions $f_{1}$, $f_{2}$ and $f_{3}$. As the latter relabeling
will have no effect on the unambiguous distinguishability of the
oracle operators, clearly, neither will the former.

This condition also gives an alternative perspective on why four
totally indistinguishable functions with $M=2$ cannot have
unambiguously distinguishable oracle operators.  For $M=2$, the
function matrix has only two columns and so not all three of the
required column types can occur.

The case of $M=3$ is also noteworthy, because here, if inequality
\eqref{niineq} is satisfied, then Eq. \eqref{summi} implies that the
values of the $N_{i}$ are uniquely specified. We must have $N_{1}=0$
and $N_{i}=1$ for $i=2,3,4$.  As a simple example, we may choose the
following four functions in ${\cal F}_{32}$:
\begin{eqnarray}
\label{idgrover}
f_{0}:(0,1,2)&{\mapsto}&(1,1,1), \\
f_{1}:(0,1,2)&{\mapsto}&(1,0,0), \\
f_{2}:(0,1,2)&{\mapsto}&(0,1,0), \\
\label{idgrover4}
f_{3}:(0,1,2)&{\mapsto}&(0,0,1).
\end{eqnarray}
It can be seen that $f_{0}$ is a uniform function and, from Eq.
\eqref{groverfunctions}, that $f_{1}$, $f_{2}$ and $f_{3}$
correspond to a three-element unsorted database search.

One additional interesting feature of this condition is that it
places no constraints on the $a_{x}$.  In other words, if we wish to
construct a set of four totally indistinguishable functions in
${\cal F}_{MN}$, then one of the functions, in our description
$f_{0}$, may be chosen arbitrarily.  The required properties of the
entire set will constrain the form of the remaining functions in
relation to $f_{0}$.  In the case of $M=3$, the freedoms we have in
defining the remaining functions are as follows.  The three columns
types are predetermined and there must be one column of each type,
although their locations may be chosen freely.  Also, the $a_{x}$
and $\bar{a}_{x}$ may be arbitrary, non-equal numbers in
$\mathbb{Z}_{N}$.

\section{Unambiguous oracle operator discrimination with multiple parallel calls}
\renewcommand{\theequation}{7.\arabic{equation}}
\setcounter{equation}{0} \label{sec:7}

So far, we have been considering unambiguous oracle operator
discrimination with only one call to the oracle.  A natural question
to ask then is, if a set of oracle operators are not unambiguously
distinguishable in this one shot scenario, can we overcome this by
making multiple calls?

In the most general scenario we can consider,  a register used for
one call to the oracle can be reused for subsequent calls.  Here, we
make the simplifying assumption that such reuse does not take place
and that instead separate oracle calls occur in parallel and upon
different registers.  However, collective measurements are assumed
to be possible on these registers.

If we can make $C$ parallel oracle calls, then the oracle operators
will be unambiguously distinguishable iff the $C$-fold tensor
products $U_{f_{j}}^{{\otimes}C}$ are linearly independent.  From
Section \ref{sec:2}, we know that we can check this by determining
whether or not the Gram matrix of these operators is non-singular.
We find that the elements of this matrix are
\begin{equation}
\mathrm{Tr}(U_{f_{j'}}^{{\dagger}{\otimes}C}U_{f_{j}}^{{\otimes}C})=\left(\mathrm{Tr}(U_{f_{j'}}^{{\dagger}}U_{f_{j}})\right)^{C}=(N{\Gamma}_{j'j})^{C}.
\end{equation}
So, neglecting the irrelevant factor of $N^{C}$, the matrix which
must be non-singular for the operators $U_{f_{j}}^{{\otimes}C}$ to
be unambiguously distinguishable has elements ${\Gamma}_{j'j}^{C}$.
We recognise this as the $C$th Hadamard (i.e. entrywise) power of
${\Gamma}$ and denote it by ${\Gamma}^{{\circ}C}$.

We note that this problem is closely related to unambiguous
discrimination among quantum states with multiple copies.  Although
one cannot unambiguously discriminate among a set of linearly
dependent pure states with just one copy, if we have $C$ copies of
the state, then these $C$-fold copy states may be linearly
independent and thus amenable to collective unambiguous
discrimination.  Upper and lower bounds upon the number of copies
required for this to be possible have been obtained in terms of the
number of states to be discriminated and the dimensionality of
subspace spanned by the possible single-copy states \cite{Clin}.
Unfortunately, we have found it impractical to apply these bounds to
unambiguous oracle operator discrimination, as to make use of them
would require us to know the dimensionality of the subspace spanned
by the set of possible oracle operators, which seems to be difficult
to determine in general.  Instead, we take an alternative approach,
using a certain result from matrix analysis, to obtain a sufficient
condition for unambiguous oracle operator distinguishability with
$C$ parallel calls.  To proceed, we use the following definition:

\begin{definition}[Diagonal and strict diagonal dominance]  A $K{\times}K$
matrix $A=(a_{j'j})$, where $j,j'=0,{\ldots},K-1$, is said to be
diagonally dominant if
\begin{equation}
|a_{jj}|{\geq}\sum_{j'=0\atop
j'{\neq}j}^{K-1}|a_{j'j}|\;\;{\forall}\;\;j=0,{\ldots},K-1.
\end{equation}
It is said to be strictly diagonally dominant if the strict
inequality holds here for all $j=0,{\ldots},K-1$.
\end{definition}
One of the key properties of strictly diagonally dominant matrices
is that they are non-singular \cite{HJ}.  We can then use the
condition for strict diagonal dominance to test for the
non-singularity of ${\Gamma}^{{\circ}C}$.  Making use of the fact
that ${\Gamma}_{jj}=M$ and that all elements of ${\Gamma}$ are real
and non-negative, we find that ${\Gamma}^{{\circ}C}$ will be
non-singular if
\begin{eqnarray}
\label{mcineq} &M^{C}&>\sum_{j'=0\atop
j'{\neq}j}^{K({\sigma})-1}{\Gamma}^{C}_{j'j}\;\;{\forall}\;\;j=0,{\ldots},K({\sigma})-1
\\ \label{mineq}{\Leftrightarrow}&M&>\left(\sum_{j'=0\atop
j'{\neq}j}^{K({\sigma})-1}{\Gamma}^{C}_{j'j}\right)^{1/C}\;\;{\forall}\;\;j=0,{\ldots},K({\sigma})-1.
\end{eqnarray}
This is a sufficient condition for unambiguous discrimination among
standard oracle operators with $C$ parallel calls.  As a simple
example of how it can be used, consider the four functions in ${\cal
F}_{22}$ shown in Eqs. \eqref{f221}-\eqref{f224} with corresponding
matrix ${\Gamma}$ given by Eq. \eqref{f22gamma}, which is singular.
We find that $\sum_{j'=0\atop
j'{\neq}j}^{K({\sigma})-1}{\Gamma}^{C}_{j'j}=2$ for all
$C{\in}\mathbb{R}$ and so \eqref{mineq} leads to the requirement
that $C>1$. This implies that although the oracle operators
corresponding to these functions cannot be unambiguously
discriminated with one call, they can with two parallel calls.  We
should expect this since we can also discriminate among the
functions in ${\cal F}_{2N}$ with two calls to a classical oracle,
by simply evaluating the function for the two possible values of
$x$.

We will now use \eqref{mcineq}, which is a set of inequalities, to
obtain a single inequality which provides a general condition
specifying a number of parallel calls $C$ which is sufficient for
unambiguous oracle operator discrimination to be possible.  Let us
define ${\delta}_{min}$ as the minimum, over all pairs of functions
in ${\sigma}{\subset}{\cal F}_{MN}$, of the number of values of $x$
for which the values of the functions in each pair differ.  Then we
find that
\begin{equation}
\sum_{j'=0\atop
j'{\neq}j}^{K({\sigma})-1}{\Gamma}^{C}_{j'j}{\leq}(K({\sigma})-1)(M-{\delta}_{min})^{C}\;\;{\forall}\;\;j=0,{\ldots},K({\sigma})-1.
\end{equation}
It follows that if
\begin{equation}
M^{C}>(K({\sigma})-1)(M-{\delta}_{min})^{C},
\end{equation}
then the condition in \eqref{mcineq} for the strict diagonal
dominance of ${\Gamma}^{{\circ}C}$ is automatically satisfied. Using
the elementary properties of logarithms, we find that this
expression is equivalent to
\begin{equation}
\label{calls}
C>\frac{{\ln}(K({\sigma})-1)}{{\ln}(M)-{\ln}(M-{\delta}_{min})}.
\end{equation}
We see that we have obtained here a sufficient condition on the
number of parallel oracle calls $C$ for unambiguous discrimination
among the standard oracle operators to be possible, in terms of
quantities which are intrinsic properties of the set of functions
${\sigma}$ itself.

\section{Interconvertibility of standard and entanglement-assisted minimal oracle operators }
\renewcommand{\theequation}{8.\arabic{equation}}
\setcounter{equation}{0} \label{sec:8}

The final topic we shall discuss is an intriguing relationship
between the standard and entanglement-assisted minimal oracle
operators for a set of permutations in ${\cal F}_{MM}$. This
relationship is a simple consequence of properties of Gram matrices
and oracle operators that have arisen earlier in this article. The
first of these is the fact that if we have two sets of vectors in
the same vector space and with the same Gram matrix, then these sets
can be unitarily transformed into each other.  We made use of this
in Section \ref{sec:2}. The second arose originally in our proof of
Theorem \ref{usdoracle}.  This is the fact that, for fixed $M$, the
standard and entanglement-assisted minimal oracle operators have the
same Gram matrices.

As we saw in Section \ref{sec:2}, it is often useful to treat
operators as vectors in a vector space.  For $N=M$, the standard
oracle operators and entanglement-assisted minimal oracle operators
are elements of ${\cal B}({\cal H}_{M}^{{\otimes}2})$, the vector
space consisting of bounded operators on ${\cal
H}_{M}^{{\otimes}2}$, with boundedness being guaranteed on a
finite-dimensional vector space. The above considerations lead to

\begin{theorem}
For each integer $M{\geq}1$ and for every permutation $f{\in}{\cal
F}_{MM}$,  there exists a single unitary operator ${\cal U}:{\cal
B}({\cal H}_{M}^{{\otimes}2}){\mapsto}{\cal B}({\cal
H}_{M}^{{\otimes}2})$ such that
\begin{equation}
{\cal U}(\bar{Q}_{f})=U_{f}
\end{equation}
where $\bar{Q}_{f}$ is the entanglement-assisted minimal oracle
operator and $U_{f}$ the standard oracle operator corresponding to
$f$.
\end{theorem}
This quite remarkable result holds true in spite of the fact,
pointed out by Kashefi et al. \cite{KKVB}, that the number of
invocations of a standard oracle operator corresponding to a
permutation $f{\in}{\cal F}_{MM}$ required to produce the
corresponding minimal oracle operator grows as $O(\sqrt{M})$.

The key to understanding this is that ${\cal U}$ does not, in
general, represent a physical transformation of the $\bar{Q}_{f}$
into the $U_{f}$ for all probe states. Being an operator on a space
of operators rather than on a space of states, ${\cal U}$ is
actually a superoperator which does not, in general, describe any
physical process enabling the entanglement-assisted minimal oracle
operators to be simulated by standard oracle operators or vice
versa.

Although the results of Kashefi et al. are sufficient to exclude the
possibility of the unitary superoperator ${\cal U}$ representing a
physical transformation in general, the conditions under which one
arbitrary set of unitary operators can be simulated by some other
are not yet known, at least not it terms which are more helpful than
the obvious requirement of the existence of appropriate completely
positive, linear, trace-preserving maps. To examine the contrast
between the unitary superoperator ${\cal U}$ and actual physical
transformations, we shall restrict the latter to be general unitary
transformations of operators on the same space as that upon which
these operators act. We take such a transformation to involve
unitary operators $S,T{\in}{\cal B}({\cal H}_{M}^{{\otimes}2})$ such
that
\begin{equation}
\label{sttrans}
S\bar{Q}_{f}T=U_{f}
\end{equation}
for every permutation $f{\in}{\cal F}_{MM}$.  We shall refer to such
a transformation as a bilateral unitary transformation.

Interestingly, for the simplest non-trivial case, which is that of
$M=2$, a bilateral unitary transformation between the two sets of
oracle operators does exist. Here, we have two permutations, the
identity function and the logical NOT operation. The standard and
entanglement-assisted minimal oracle operators for these functions
may be written in terms of the Pauli spin operators as
\begin{eqnarray}
U_{\mathrm{ID}}&=&|0{\rangle}{\langle}0|{\otimes}{\id}_{2}+|1{\rangle}{\langle}1|{\otimes}{\sigma}_{x}=\mathrm{CNOT}, \\
U_{\mathrm{NOT}}&=&|0{\rangle}{\langle}0|{\otimes}{\sigma}_{x}+|1{\rangle}{\langle}1|{\otimes}{\id}_{2}, \\
\bar{Q}_{\mathrm{ID}}&=&{\id}_{2}{\otimes}{\id}_{2}, \\
\bar{Q}_{\mathrm{NOT}}&=&{\sigma}_{x}{\otimes}{\id}_{2},
\end{eqnarray}
where $|0{\rangle}$ and $|1{\rangle}$ are the eigenstates of
${\sigma}_{z}$ with eigenvalues $+1$ and $-1$ respectively and
${\id}_{2}$ is the identity operator on the Hilbert space of a
qubit. Suitable unitary operators $S$ and $T$ for carrying out the
transformation in Eq. \eqref{sttrans} are
\begin{eqnarray}
S&=&({\id}_{2}{\otimes}P_{+}+i{\sigma}_{z}{\otimes}P_{-})\mathrm{SWAP}, \\
T&=&\mathrm{SWAP}({\id}_{2}{\otimes}(P_{+}-iP_{-})),
\end{eqnarray}
where $P_{\pm}$ are the projectors onto the eigenstates of
${\sigma}_{x}$ with eigenvalues ${\pm}1$. So, we see that for $M=2$,
we can indeed have a bilateral unitary transformation between the
standard oracle operators and the entanglement-assisted minimal
oracle operators. However, this is not possible for any $M>2$.

To see why not, we note that we can eliminate $T$ in the following
way. For any $M$, we have $\bar{Q}_{ID}={\id}_{M}^{{\otimes}2}$,
from which Eq. \eqref{sttrans} gives $ST=U_{ID}$, implying that
$T=S^{\dagger}U_{ID}$.  Substituting this into Eq. \eqref{sttrans}
gives the equivalent single unitary operator transformation
\begin{equation}
\label{strans} S\bar{Q}_{f}S^{\dagger}=U_{f}U_{ID}.
\end{equation}
Hence, we obtain
\begin{equation}
S[\bar{Q}_{f'},\bar{Q}_{f}]S^{\dagger}=[U_{f'}U_{ID},U_{f}U_{ID}],
\end{equation}
for any two permutations $f$ and $f'$ in the permutation group of
degree $M$.   For $M{\geq}3$, there exist permutations $f$ and $f'$
for which the left hand side of this expression is non-zero, because
the permutation group of degree $M$ is non-Abelian for all
$M{\geq}3$.
 However, the right hand side commutator always vanishes because the
standard oracle operators form an Abelian group.  It follows that
for permutations that do not commute, there is no bilateral unitary
transformation from the entanglement-assisted minimal oracle
operators into the standard oracle operators.

There is one limited sense, however, in which the identicality of
the Gram matrices of the standard and entanglement-assisted minimal
oracle operators does correspond to a physical process for all $M$.
Suppose that we have two pairs of $M$-dimensional quantum systems,
where each pair is a copy of the entire register upon which these
oracle operators act.  Then consider a state $|{\Phi}{\rangle}$
which is a normalised, maximally entangled state of these pairs. Let
us now, as before, index the functions in our required set
${\sigma}{\subset}{\cal F}_{MM}$ by $j$.  Here, ${\sigma}$ is the
set of permutations in ${\cal F}_{MM}$ and so $j=0,{\ldots},M!-1$.
The state which results from the action of the standard oracle
operator $U_{f_{j}}$ corresponding to the function
$f_{j}{\in}{\sigma}$, upon half of the state $|{\Phi}{\rangle}$,
will be denoted by
\begin{equation}
|U_{f_{j}}{\rangle}=(U_{f_{j}}{\otimes}{\id}_{M^{2}})|{\Phi}{\rangle},
\end{equation}
where ${\id}_{M^{2}}$ is the identity operator on ${\cal
H}_{M}^{{\otimes}2}$.  Similarly, for the entanglement-assisted
minimal oracle operators, we write
\begin{equation}
|\bar{Q}_{f_{j}}{\rangle}=(\bar{Q}_{f_{j}}{\otimes}{\id}_{M^{2}})|{\Phi}{\rangle}.
\end{equation}
It is a simple matter to show that the above sets of states have the
same Gram matrix, whose elements are given by
\begin{equation}
\label{uqgram}
{\langle}U_{f_{j'}}|U_{f_{j}}{\rangle}=\frac{1}{M^{2}}\mathrm{Tr}(U^{\dagger}_{f_{j'}}U_{f_{j}})=\frac{1}{M^{2}}\mathrm{Tr}(\bar{Q}^{\dagger}_{f_{j'}}\bar{Q}_{f_{j}})={\langle}\bar{Q}_{f_{j'}}|\bar{Q}_{f_{j}}{\rangle}=\frac{1}{M}{\Gamma}_{j'j}.
\end{equation}
It follows that the $|U_{f_{j}}{\rangle}$ and the
$|\bar{Q}_{f_{j}}{\rangle}$  are interconvertible by a physical
unitary transformation on ${\cal H}_{M}^{{\otimes}4}$. Recalling the
discussion of Theorem \ref{usdoracle}, we are rapidly led to
conclude that for any probe state $|{\Phi}{\rangle}$ of the above
form, the output states for both sets of oracle operators are
equally distinguishable for any discrimination strategy. So in this
sense, the equality of the Gram matrices of both types of oracle
operator does have an operational interpretation.  Indeed, Eq.
\eqref{uqgram} may serve to suggest a further interpretation of the
matrix ${\Gamma}$ itself, where it appears as being equal, up to a
proportionality factor, to the Gram matrix of the states produced by
either kind of oracle operator for a probe state $|{\Phi}{\rangle}$
of the form we have described. Nevertheless, the fact that we
require a specific kind of probe state implies that this does not
lead to any general conclusions relating to the comparison of the
distinguishability properties of both types of oracle operator.

\section{Discussion}
\renewcommand{\theequation}{9.\arabic{equation}}
\setcounter{equation}{0} \label{sec:9}

The aim of the present article has been to investigate the
possibility of unambiguous discrimination among oracle operators.
Our motivation for this comes primarily from quantum computation,
where the oracle identification problem plays a key role.  In most
existing treatments of this problem, the measurement which is used
to identify the oracle operator is taken to be a simple projective
measurement. Unambiguous measurements are more powerful and allow to
us discriminate in an error-free manner among non-orthogonal states.
As a result, a considerable amount of attention has been given to
them in recent years. The basic theory of unambiguous state
discrimination is now highly-developed
\cite{discreview1,discreview2} and such measurements have been
frequently applied to problems in quantum cryptography
\cite{Norbert,Trine,SARG,HSFK}.  The related problem of unitary
operator discrimination, which has been our main concern here, is
also beginning to play an important role in this field
\cite{Long2,Long3}.  As such, an interesting question is whether or
not such measurements have a similarly useful role
 to play in relation to the other main aspect of applied quantum information science, which is quantum computation.  Since the acquisition
 of classical information during, or at the end, of a quantum
 computation often takes place as the result of an oracle
 query, the possibility of unambiguous discrimination among oracle
 operators seems to be the most natural place to start investigating
 the applicability of this type of measurement to this field.

 Our emphasis has not been on the details of the
 measurements required to perform unambiguous oracle operator
 discrimination.  These are unambiguous state
 discriminating measurements tailored to the particular set of oracle operators and to
 the
 probe state which has been prepared.  As such, one can apply the
 numerous results already established in relation to the
 construction of these measurements \cite{SHB,MSB,HB}.   However, in the context of quantum computation, it would be desirable to have an understanding of the
 complexity of such measurements.  Here, we have
 focused
 mainly on the problem of determining whether or not a given set of
oracle
 operators can be unambiguously discriminated with some such measurement.  Logically, this is the most fundamental problem in relation to this topic.  However, as we hope to have
 demonstrated in this article, it is extremely
 rich and its solutions for various cases yield new insights into,
 for example,
 existing quantum algorithms, such as in our discussion of
 unambiguous discrimination among the Grover oracle operators.

This article only serves as an initial investigation into
unambiguous oracle operator discrimination. There are undoubtedly
intriguing new things to be discovered in relation to the problem of
determining whether or not a given set of oracle operators are
unambiguously distinguishable. Of particular significance are
situations where such discrimination represents a non-classical
effect.  In Section \ref{sec:6}, we considered unambiguous oracle
operator discrimination where the corresponding functions possess
the property of total indistinguishability, i.e. they can never be
discriminated classically.  We obtained some quite general results
in relation to this matter.  Two of these were constraints, namely
the simple fact that there must be at least four functions in a
totally indistinguishable set of distinct functions and that for a
finite set of such functions on a Boolean domain, the standard
oracle operators are never unambiguously distinguishable. We then
gave a complete description of sets of four totally
indistinguishable functions with unambiguously distinguishable
standard oracle operators.

One point that should be made about the latter two results is that
although we took the domain and range of the set of functions to be
$\mathbb{Z}_{M}$ and $\mathbb{Z}_{N}$ respectively, one can easily
verify that the proofs are somewhat insensitive to this.  We can,
for example, straightforwardly generalise the domain and range to
sets of arbitrary, finite, complex numbers whose cardinalities are
the same as the original integer sets with the main conclusions
unchanged. This generalisation is essentially minor. There are,
however, significant, non-trivial open problems in relation to such
sets of functions and their corresponding oracle operators.

A natural one to pose is: can a complete characterisation of such
sets of functions, such as we performed for those with cardinality
four in Section VI.C, be carried out for larger numbers of
functions? We expect that in general, the function matrix, in
particular the frequencies with which certain column types occur,
will play the same, important role that it did in our analysis of
the case of four functions. This seems to be assured by the fact
that the matrix ${\Gamma}$ is constructed by counting coincidences
in these columns. However, for larger numbers of functions, there is
the inevitable problem of obtaining analytically the determinants of
high dimensional matrices and being able to make general statements
about classes of such determinants. A further complicating factor
when considering larger sets of totally indistinguishable functions
is the apparent need to obtain a description of all possible column
types. In the four function case, the Boolean nature of the elements
of these columns makes this straightforward.  This property can also
be seen to hold for five functions.  However, we do not have this
luxury for larger sets of functions.

A further issue to address is the potential applicability of sets of
totally distinguishable functions with unambiguously distinguishable
oracle operators.   Can this intriguing, non-classical effect serve
as the basis for novel quantum protocols?  The simple example we
gave in Eqs. \eqref{idgrover}-\eqref{idgrover4} relates to
searching. However, we suspect that the scope for applications of
such sets of functions extends far beyond this and deserves to be
explored.

In this article, we have considered many issues which relate to, or
ensue from the unambiguous oracle operator discrimination condition.
However, there are a large number of questions that we have either
not, or have barely addressed.  Principal among these, we believe,
is the problem of optimal unambiguous oracle operator
discrimination. If it is possible to unambiguously discriminate
among a particular set of oracle operators, then what is the maximum
probability of success?  Indeed, how do the distinguishability
properties of the standard and minimal oracle operators compare? For
$M=2$,  the standard and entanglement-assisted minimal oracle
operators corresponding to permutations are related by a bilateral
unitary transformation and so we can see that in this case, both
sets of operators are equally distinguishable. However, such
transformations are not possible for $M{\geq}3$. Indeed here, the
minimal oracle operators do not mutually commute and so Theorem
\ref{commute} does not apply to them.  There is then the possibility
that optimal unambiguous discrimination among a set of such
operators requires the use of an entangled probe state. This may be
of some relevance to what we regard as being the main question here,
which is: for a given set of permutations, which kind of oracle
operators, the standard or entanglement-assisted minimal oracle
operators, have the higher unambiguous discrimination success
probability? Indeed, for any set of permutations in ${\cal F}_{MM}$
and for any integer $M{\geq}3$, if the oracle operators are
unambiguously distinguishable, then are the entanglement-assisted
minimal oracle operators always more distinguishable than the
standard oracle operators, or perhaps vice versa?

A further open problem is whether or not the framework we have
developed in this article can be generalised in a simple and useful
way to unambiguous discrimination among sets of oracle operators.
As we described in the Introduction, many important quantum
algorithms involve discrimination among sets of oracle operators,
rather than fine-grained discrimination among the oracle operators
themselves. As such, there may exist circumstances where we have a
set of oracle operators which are not individually unambiguously
distinguishable, but where we only require that certain subsets of
this total set can be unambiguously discriminated from each other.
When this is the case, we are not actually interested in unambiguous
discrimination among the individual oracle operators.  Rather, we
are concerned with unambiguous discrimination among more general
quantum operations, where each operation is a mixture of the oracle
operators in each subset.  It is possible that the recent results of
Wang and Ying \cite{WY} relating to unambiguous discrimination among
general quantum operations are applicable to this problem.

Finally, we shall describe what we regard as being the most pressing
open questions concerning unambiguous oracle operator discrimination
with multiple calls.  Although this article has focused mainly on a
single call to the oracle, in Section \ref{sec:7}, we did consider
multiple parallel calls and obtained a sufficient condition
\eqref{calls} on the number of such calls to enable unambiguous
standard oracle operator discrimination for a given set of
functions.  Is it possible to move forward in this direction by, for
example, providing a tighter sufficient condition and/or a suitably
non-trivial necessary condition?  There is also the obvious
generalisation to non-parallel calls to be considered.  This leads
us to what we may term the unambiguous query complexity problem,
which we may state in the following way: for a given set of
functions ${\sigma}{\in}{\cal F}_{MN}$, how many uses of the
standard oracle operators, interspersed with arbitrary unitary
operators and making use of ancillas, are sufficient to produce a
set of linearly independent output states for some probe state? This
is equivalent to the requirement that the corresponding products of
multiple oracle and arbitrary unitary operators, with the latter
being independent of the functions under consideration, are linearly
independent.

To conclude, unambiguous discrimination among oracle operators is an
important potential application of unambiguous discrimination
measurements to quantum computation. In this article, we have laid
the foundations for the further exploration of this possibility.
However, much progress remains to be made before we have a full
understanding of the scope and limitations of unambiguous
discrimination within this context.

\section*{Acknowledgements}
AC would like to thank Timothy Spiller for helpful and interesting
discussions.  He would also like to thank Richard Jozsa for this
reason and for pointing out ref. \cite{BBHT}. AC was supported by
the EU project QAP.

\section*{Appendix: Proof of Theorem \ref{cycle}}
\renewcommand{\theequation}{A.\arabic{equation}}
\setcounter{equation}{0} \label{app}

Here, we provide a proof of Theorem \ref{cycle} from Section
\ref{sec:6}, which we
restate here for convenience:\\

{\noindent \em Every connected component of a graph $G({\sigma})$,
where ${\sigma}{\subset}{\cal F}_{2N}$ is a finite set of totally
indistinguishable functions, has an induced
subgraph which is an even cycle of length ${\geq}4$.}\\

\noindent{\bf Proof}: The following proof is constructive.  We give
a procedure for constructing a certain kind of cycle which we then
show has all of the desired properties:
\\

\noindent (1) Let us begin with an arbitrary vertex in $G({\sigma})$
and denote it by $V_{j_{1}}$. We start to construct a graph
$G({\sigma}')$, where ${\sigma}'$ is initially the empty set, by
adding $f_{j_{1}}$ to ${\sigma}'$ and therefore $V_{j_{1}}$ to
$G({\sigma}')$.  \\

\noindent (2) We now choose an arbitrary vertex in $G({\sigma})$
which is $X$-adjacent to $V_{j_{1}}$, denoting it by $V_{j_{2}}$. We
add $f_{j_{2}}$ to ${\sigma}'$.  We add both this vertex and the
edge linking it to $V_{j_{1}}$  to
$G({\sigma}')$.\\

\noindent (3) We now choose an arbitrary vertex in $G({\sigma})$
which is $Y$-adjacent to $V_{j_{2}}$, denoting it by $V_{j_{3}}$. We
add $f_{j_{3}}$ to ${\sigma}'$.  As above, we add both this vertex
and the edge linking it to $V_{j_{2}}$ to
$G({\sigma}')$.\\

\noindent (4) We keep repeating the above two steps until a certain
condition, which we specify in step (5), is satisfied.  This
repetition means that for each odd $r$, we add $f_{j_{r+1}}$ to
${\sigma}'$, where $f_{j_{r+1}}$ is any function in ${\sigma}$ whose
corresponding vertex $V_{j_{r+1}}$ is $X$-adjacent in $G({\sigma})$
to $V_{j_{r}}$. We also add both the vertex $V_{j_{r+1}}$ and the
edge linking it to $V_{j_{r}}$ to $G({\sigma}')$.  In the case of
even $r$, we add $f_{j_{r+1}}$ to ${\sigma}'$, where $f_{j_{r+1}}$
is any function in ${\sigma}$ whose corresponding vertex is
$Y$-adjacent in $G({\sigma})$ to $V_{j_{r}}$. We also add both this
vertex $V_{j_{r+1}}$ and the edge linking it to $V_{j_{r}}$
to $G({\sigma}')$.\\

\noindent (5) We terminate this repetition after we have added the
first vertex we can which is adjacent in $G({\sigma})$ to a previous
vertex $V_{j_{r'}}$ for some $r'<r-1$.  We denote this particular
value of $r$ by $R$.  When we reach $V_{j_{R}}$, there may be
several vertices $V_{j_{r'}}$ we can choose among. When this is so,
we choose the one with the largest value of $r'$, which we denote by
$R'$.  We complete a cycle by adding to $G({\sigma}')$ the edge
linking vertices
$V_{j_{R}}$ and $V_{j_{R'}}$.\\

\noindent (6)   We delete from $G({\sigma}')$ all vertices
$V_{j_{r}}$ for $r<R'$ and all edges attached to these vertices. The
resulting graph, our final $G({\sigma}')$, having vertices
$V_{j_{R'}},{\ldots},V_{j_{R}}$, is a subgraph of $G({\sigma})$ with
all of the desired properties, as we
shall now prove. \\

\noindent {\em Existence of cycle}:  The fact that, for a finite set
of functions, we are indeed able to construct a cycle this way, i.e.
the inevitability of step (5) taking effect, can be seen in the
following way.  We know that for any vertex $V{\in}G({\sigma})$,
there exists at least one vertex in $G({\sigma})$ which is
$X$-adjacent to $V$ and at least one which is $Y$-adjacent to $V$.
These two vertices are, of course, different; otherwise, they would
both be $V$ itself. It follows that, were it not for the termination
condition specified in step (5), we could endlessly repeat step (4).
However, on a finite graph, this repetition will inevitably revisit
vertices previously incorporated into $G({\sigma}')$.  It is when
this is about to happen that the termination condition is triggered.
There is no freedom in choosing the final edge to be incorporated
into $G({\sigma}')$, as this is done in the way which makes the
shortest possible cycle. Following this, the final deletion step (6)
removes
all vertices and edges which are not part of this cycle.  \\

\noindent {\em Even length}:  That the cycle has even length can be
seen from the fact that, up until step (5), the graph $G({\sigma}')$
is constructed by adding to it alternating horizontal and vertical
edges together with corresponding vertices. From this, we see that
the only way in which the cycle could have odd length would be if
this alternation were suspended at the closure step (5), which would
result in three vertices, two of which are $V_{j_{R}}$ and
$V_{j_{R'}}$, being mutually $X$- or $Y$-adjacent, i.e. collinear.
Since $V_{j_{R'}}$ immediately follows $V_{j_{R}}$, the third
collinear vertex would have to come before $V_{j_{R}}$ or after
$V_{j_{R'}}$, in the sense of direction of the cycle.  The first
possibility contradicts the fact that we terminated the repetition
step at the earliest possible opportunity, since it would have
allowed us to perform the termination at the preceding vertex.  The
second possibility is inconsistent with the fact that $V_{j_{R}}$ is
the furthest vertex along the cycle which is adjacent to
$V_{j_{R'}}$ in $G({\sigma})$, since we could have chosen the next
vertex instead.
Therefore, the cycle we have constructed has even length.\\

\noindent {\em Length ${\geq}4$}: That this cycle has length
${\geq}4$ can be established in the following way.  From step (5),
we have $R'<R-1$ and so there are at least 3 vertices in the final
cycle.  Having established that this cycle is even, we see that it
must
therefore be of length ${\geq}4$.\\

\begin{figure}
\begin{center}
\epsfxsize16cm \centerline{\epsfbox{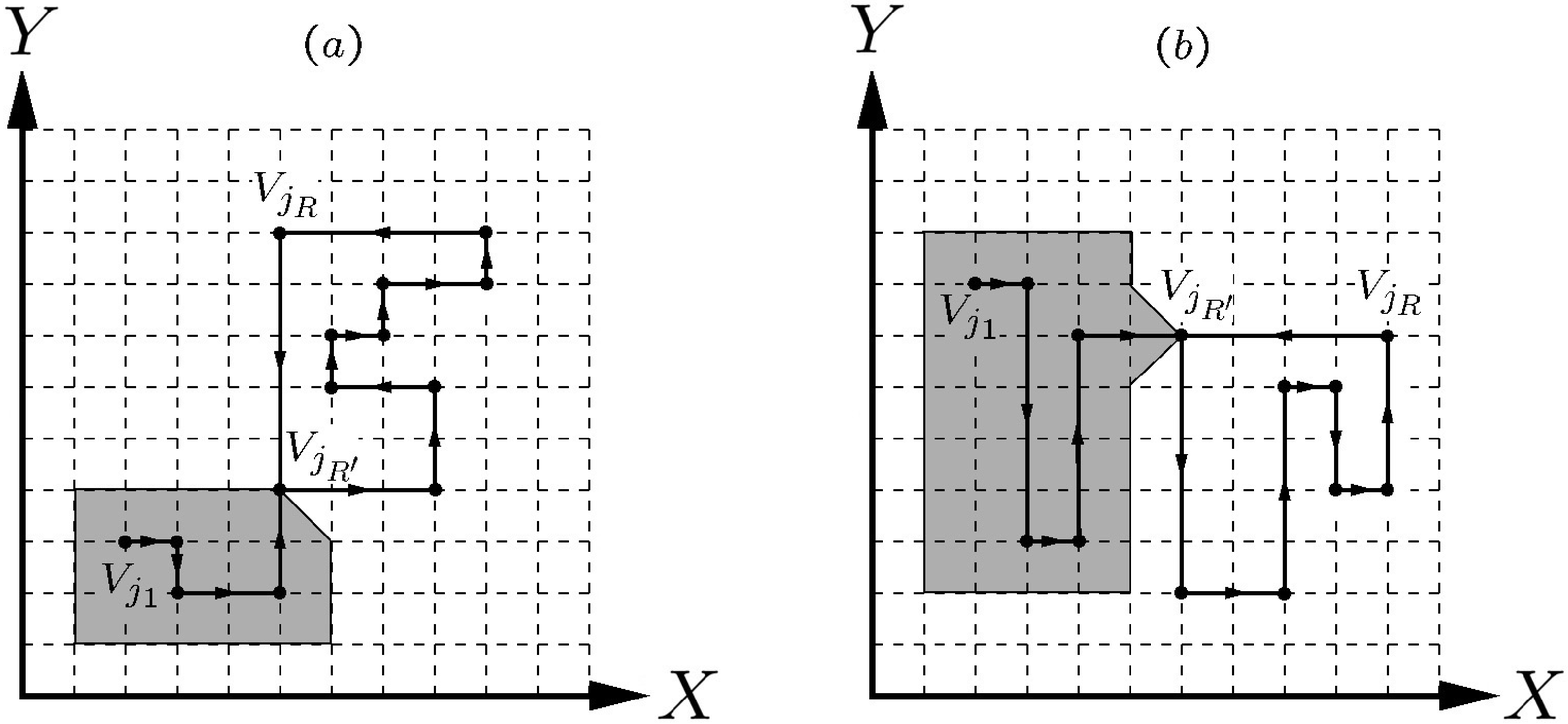}}
\end{center}
\caption{Construction of the final cyclic subgraph $G({\sigma}')$ of
$G({\sigma})$ with significant vertices indicated.  The two cases
(a) and (b) indicate the two ways in which the cycle can be closed.
The shaded regions contain the vertices and edges to be deleted in
order to obtain the final $G({\sigma}')$.} \label{figure3}
\end{figure}

\noindent {\em Induced subgraph of $G({\sigma})$}: If the final
cycle were not an induced subgraph of $G({\sigma})$, then there
would be further edges linking vertices in the cycle set
$\{V_{j_{R'}},{\ldots},V_{j_{R}}\}$ to each other in the original
graph $G({\sigma})$, i.e. in addition to those which form part of
the cycle graph $G({\sigma}')$ itself. Suppose that this were the
case, that is, that there existed vertices $V_{j_{R_{0}}},
V_{j_{R_{1}}}{\in}G({\sigma}),G({\sigma}')$ with this property.
Without loss of generality, we may take $R_{1}>R_{0}$.  Indeed, by
assumption these vertices are not adjacent in $G({\sigma}')$, so we
may take $R_{1}>R_{0}+1$.  They are, however, adjacent in
$G({\sigma})$.  This implies that during the construction of
$G({\sigma}')$, we would have been obliged to terminate the
repetition of (4) on encountering $V_{j_{R_{1}}}$ and obtain
$R_{1}=R$. There are two possibilities for what would happen next.
Either, on completion of step (5), this vertex would have been made
adjacent to $V_{j_{R_{0}}}$ in $G({\sigma}')$ and we would have
$R_{0}=R'$, contradicting the assumption that these two vertices are
not adjacent in this graph, or $R_{0}<R'$.  In the latter scenario,
the vertex $V_{j_{R_{0}}}$ would be removed from $G({\sigma}')$ in
step (6), contradicting the premise that $V_{j_{R_{0}}}$ belongs to
this final cycle graph. This shows that, indeed, the final
$G({\sigma}')$ is an induced subgraph of $G({\sigma})$ as desired.\\

\noindent {\em Occurrence within an arbitrary, connected component
of $G({\sigma})$}: Finally, we note that our starting vertex
$V_{j_{1}}$ is an arbitrary vertex in $G({\sigma})$.  This, together
with the fact that, at every stage of its construction,
$G({\sigma}')$ is connected, implies that every connected component
of $G({\sigma})$ has a cyclic subgraph of the form we have
described. This completes the proof.$\Box$\\

Figure \ref{figure3} depicts the construction of the final cyclic
subgraph $G({\sigma}')$.  Due to the time-dependent nature of our
procedure, it is convenient, for the purposes of exposition, to
regard $G({\sigma}')$ as a directed graph during its construction.
The direction of each edge indicates the location of the next vertex
from $G({\sigma})$ to be incorporated into $G({\sigma}')$ in steps
(2)-(4) above.  Of course, this edge itself is also incorporated
into $G({\sigma}')$.  The two situations illustrated, denoted by (a)
and (b), correspond to closure of the cycle in step (5) along either
the horizontal or vertical axis. In each case, the shaded region
contains the vertices and edges to be deleted in step (6) in order
to obtain the final cyclic subgraph.


\begin{thebibliography}{99}
\bibitem{ring}
$\mathbb{Z}_{M}$ is the set of integers from 0 to $M-1$, inclusive,
likewise with $\mathbb{Z}_{N}$.
\bibitem{ccomp}
An induced subgraph of a graph $G$ is a subset of the vertices of
$G$ together with all edges in $G$ which link vertices in this
subset to each other. A connected component of a graph $G$ is an
induced, connected subgraph of $G$ whose vertices are disconnected
from all other vertices in $G$.
\end{thebibliography}
\end{document}